\documentclass{emulateapj}
\slugcomment{ApJ}

\shorttitle{RLAGN are mergers}
\shortauthors{Chiaberge et al.}

\begin{document}

%% LaTeX will automatically break titles if they run longer than
%% one line. However, you may use \\ to force a line break if
%% you desire.

\title{Radio Loud AGN are Mergers}

%% Use \author, \affil, and the \and command to format
%% author and affiliation information.
%% Note that \email has replaced the old \authoremail command
%% from AASTeX v4.0. You can use \email to mark an email address
%% anywhere in the paper, not just in the front matter.
%% As in the title, use \\ to force line breaks.

\author{Marco Chiaberge\altaffilmark{1,2,3}, Roberto Gilli\altaffilmark{4}, Jennifer Lotz\altaffilmark{1}, 
and Colin Norman\altaffilmark{1,5} }

\altaffiltext{1}{Space Telescope Science Institute, 3700 San Martin Drive, Baltimore, MD 21218}
\altaffiltext{2}{INAF - IRA, Via P. Gobetti 101, Bologna, I-40129}

\altaffiltext{4}{INAF – Osservatorio Astronomico di Bologna, via Ranzani 1, 40127 Bologna, Italy}
\altaffiltext{3}{Center for Astrophysical Sciences, Johns Hopkins University,
3400 N. Charles Street, Baltimore, MD 21218, USA}
\altaffiltext{5}{Department of Physics and Astronomy, Johns Hopkins University, Baltimore, MD 21218, USA}

%% Notice that each of these authors has alternate affiliations, which
%% are identified by the \altaffilmark after each name.  Specify alternate
%% affiliation information with \altaffiltext, with one command per each
%% affiliation.

%% Mark off your abstract in the ``abstract'' environment. In the manuscript
%% style, abstract will output a Received/Accepted line after the
%% title and affiliation information. No date will appear since the author
%% does not have this information. The dates will be filled in by the
%% editorial office after submission.

\begin{abstract}
We measure the merger fraction of Type~2 
radio-loud and radio--quiet 
active galactic nuclei at $z>1$ using new samples. 
The objects have HST images taken with WFC3 in the
IR channel.  These samples are compared to the 3CR sample of 
radio galaxies at $z>1$ and to a sample of non-active galaxies. We also consider
lower redshift radio galaxies with HST observations and previous generation instruments (NICMOS and
WFPC2). The full sample spans an unprecedented range in both redshift and AGN luminosity.
We perform statistical tests to determine whether the different samples are differently associated with
mergers. We find that all (92\%$^{+8\%}_{-14\%}$)  radio-loud  galaxies at $z>1$ are associated 
with recent or ongoing merger events. Among the radio-loud
population there is no evidence for any dependence of the merger fraction on either redshift or AGN power.
For the matched radio-quiet samples,
only 38\%$^{+16}_{-15}$ are merging systems.  The merger fraction for
the sample of non-active galaxies at $z>1$ is indistinguishable from radio-quiet objects. This
is strong evidence that mergers are the triggering mechanism for the 
radio-loud AGN phenomenon and the launching of relativistic jets from supermassive black holes. 
We speculate that major BH-BH mergers play a major role in spinning up the central 
supermassive black holes in these objects.

\end{abstract}

%% Keywords should appear after the \end{abstract} command. The uncommented
%% example has been keyed in ApJ style. See the instructions to authors
%% for the journal to which you are submitting your paper to determine
%% what keyword punctuation is appropriate.


%% From the front matter, we move on to the body of the paper.
%% In the first two sections, notice the use of the natbib \citep
%% and \citet commands to identify citations.  The citations are
%% tied to the reference list via symbolic KEYs. The KEY corresponds
%% to the KEY in the \bibitem in the reference list below. We have
%% chosen the first three characters of the first author's name plus
%% the last two numeral of the year of publication as our KEY for
%% each reference.

%% Authors who wish to have the most important objects in their paper
%% linked in the electronic edition to a data center may do so by tagging
%% their objects with \objectname{} or \object{}.  Each macro takes the
%% object name as its required argument. The optional, square-bracket 
%% argument should be used in cases where the data center identification
%% differs from what is to be printed in the paper.  The text appearing 
%% in curly braces is what will appear in print in the published paper. 
%% If the object name is recognized by the data centers, it will be linked
%% in the electronic edition to the object data available at the data centers  
%%
%% Note that for sources with brackets in their names, e.g. [WEG2004] 14h-090,
%% the brackets must be escaped with backslashes when used in the first
%% square-bracket argument, for instance, \object[\[WEG2004\] 14h-090]{90}).
%%  Otherwise, LaTeX will issue an error. 

\section{Introduction}

One of the most important issues in modern astrophysics is understanding the co-evolution
of galaxies and their central supermassive black holes (SMBH) 
\citep[][for recent reviews on the subject]{heckmanbest14,alexanderhickox14}. 
Both numerical simulations
and theoretical arguments show that black hole (BH) growth occurs during short-lived periods ($\sim 10^7-10^8$yr) of
intensive accretion which are also associated with powerful {\it quasar} activity 
\citep[]{soltan82,rees84,dimatteo05,hopkins08,somerville08}.
These are also
expected to correspond to periods in which galaxies grow hierarchically. Since the matter that
ultimately accretes onto the central black hole needs to lose almost all ($\sim 99.9\%$) of
its angular momentum, studies of mergers, tidal interactions, stellar bars and disk
instabilities are central for understanding the details of such a process.
Numerical simulations and analytic calculations have shown that major (gas-rich) 
mergers are capable of efficiently driving gas inflows 
towards the central region of the galaxy ($\lesssim 1kpc$) 
through tidal forces \citep[e.g.][]{hernquist89,dimatteo05,lihernquist07,hopkinsquataert11}, 
and ultimately drive the gas to the 
central $\sim 1pc$, forming an accretion disk around the central SMBH \citep[]{hopkinsquataert10}.  
This is also thought to be a likely scenario for the formation of galaxy spheroids \citep[e.g.][]{hopkins08b}.
However, disk instabilities and minor mergers may also be able to provide material for black hole accretion 
\citep[e.g.][]{hernquistmihos95,menci14}. 

It is extremely important to study the connection between galaxy and black hole growth 
from a purely observational point of view. One of the central questions is 
whether mergers or  other mechanisms may constitute the main triggering mechanisms for active galactic nuclei (AGN).
A number of recent papers investigated this issue using data from different surveys of galaxies and AGNs.
Results are often contradictory. Although it is known that not all of them are AGNs, powerful Ultra--Luminous 
Infrared Galaxies (ULIRG) 
are ubiquitously associated with major mergers \citep[]{sandersmirabel96,veilleux02}. 
According to some models \citep[e.g][]{hopkins08,hopkins08b} these objects are believed to represent a 
fundamental stage in the process of the formation of elliptical galaxies. \citet[]{kartaltepe12} showed that for 
a sample of $z\sim 2$ ULIRGs in the GOODS-South field \citep[]{giavalisco04}, the fraction of mergers 
is up to $\sim 70\%$. For lower luminosities objects (LIRGs) in the same redshift range
the merger fraction found by the same authors is significantly smaller ($\sim 30\%$).

While the merger--ULIRG connection seems to be well established, 
whether this is a viable scenario for all AGNs is still an unanswered question.
\citet[]{bahcall97} observed a sample of 20 relatively nearby QSOs with the Hubble Space Telescope (HST) 
and found that the majority of them  reside in merging galaxies. 
\citet[]{grogin03} studied the optical counterparts of the X-ray selected AGNs from the 1Msec
Chandra Deep Field South \citep[]{giacconi02}. Based on both the asymmetry index and the frequency of close companions, 
these authors concluded that mergers and interactions are not good indicators of AGN activity.
Interestingly, using HST images taken with the Wide Field Camera 3 (WFC3) 
\citet{schawinski12} showed that only a small fraction of heavily obscured QSO hosts are 
associated with mergers. \citet[]{treister12} put together a relatively large sample of literature data and
found clues for a luminosity dependence of the merger fraction in both ULIRGs and AGNs.
On the other hand, \citet[]{villforth14} showed
that  their sample of QSOs observed with HST did not show any evidence for a dependence of the merger 
fraction on luminosity. Recently, based on
a sample of SDSS galaxies, \citet{sabater15}
found that the effect of interactions is minimal in triggering AGN activity.

A major issue is related to the so-called radio-loud/radio-quiet dichotomy of active nuclei.
It has been long argued that in order to produce powerful relativistic jets, radio-loud AGNs must
possess an extra source of energy with respect to radio--quiet AGNs.
The most popular scenarios among those proposed so far assume that the energy may be extracted 
from the innermost region of a magnetized accretion disk around a rapidly spinning black hole 
\citet[]{blandfordznajek}. In the Blandford-Znajek framework, the radio-quiet/radio-loud dichotomy can
be explained in terms of a corresponding low/high black hole spin separation \citep{blandfordsaasfee}. 
Alternative models predict that jets are directly powered by the accretion disk 
\citep{blandfordpayne,pudritznorman,xu99}.
Both theoretical arguments and observational evidence support the Blandford-Znajek  scenario 
\citep[see e.g.][]{ggnature}. Recent numerical simulations also seem to confirm such a mechanism as a viable physical
explanation for efficient jet production \citep[e.g.][]{hawleykrolik06,mckinney04,tchekhovskoy11,sadowski15}.
\citet{wilsoncolbert} originally proposed that radio--loud AGNs are associated with rapidly spinning black holes
that are ultimately spun-up by major BH-BH mergers. It is clear that a straightforward prediction of such a scenario
is that radio galaxies and radio-loud QSOs (RLQSOs) should exhibit signatures of major galaxy mergers.

Radio--loud AGNs are known to be hosted by large elliptical galaxies and are associated with SMBH of at least 
$\sim 10^8$ M$_\sun$ \citep[e.g.][]{laor00,dunlop03,best05,chiabmarconi}.
Furthermore, these objects are present in richer environments (clusters and groups of galaxies) than
radio quiet AGNs \citep{shen09,donoso10,ramosalmeida13} and they are often associated with brightest 
cluster galaxies \citep{best07}.
At least a fraction of them have been known to be associated with merging system for a long time \citep[e.g.][]{heckman86,colina95}.
Recently, \citet{ramosalmeida12} studied samples of relatively low-redshift
($z<0.7$) sources with deep ground based observations. 
These authors found that the large majority ($\sim 80\%$) of radio-loud objects show disturbed morphologies. 
The same group also studied a small sample of radio--quiet Type~2 quasars \citep[]{bessiere12} and found that 
the merger fraction sample is as high as 75\%. 

However, an accurate census of the merger fraction in carefully selected samples of AGNs over 
a large range of redshift, luminosity and radio--loudness is
still needed to provide a final answer to the above questions. This should also be based
on a homogeneous set of deep, high spatial resolution observations and 
supported by firm statistical evidence for any difference in the observed merger fractions amongst the different
samples.  The aim of this paper is to
make a significant step towards such a  goal.

With the  aim of determining  the importance of mergers  in triggering
different types of AGN activity,  we select samples of both radio--loud
and radio--quiet Type~2 AGN. In this paper we focus  on Type~2 objects
only, because the bright nuclear component that dominates the emission
on Type~1  AGNs (QSOs) hampers a detailed  morphological study of both
the host  galaxies and  the close environment  of those  objects. This
constitutes  a  serious  concern,  particularly at  moderate  to  high
redshifts, even using HST images. In this work we focus on objects with
$1<z<2.5$.  This is  a range  of redshift  where the  peak of  the AGN
activity is believed to occur,  and where mergers play a dramatic role
in      the      late      evolution     of      massive      galaxies
\citep[e.g.][]{hopkins08}.  The   availability  of  HST   data  for  a
substantial number of  objects in each sample is key  to this work. In
particular,  we  focus  on  samples  observed with  WFC3  and  the  IR
channel. The extremely high  sensitivity, low background level and the
range  of wavelengths  covered by  such an  instrument  is particularly
suitable   to  our  goals.   In  fact,   the  rest   frame  wavelength
corresponding to the most widely used WFC3--IR filters (i.e. F140W and
F160W) is still well within the  optical range for objects in the redshift
range of our interest. Even  in relatively short exposures (i.e. a few
hundreds seconds,  or less  than 1 HST  orbit) the effects  of merger
events on the  structure of galaxies are clearly  revealed 
in WFC3--IR images. This is possible thanks to the high sensitivity in the near IR
coupled with the large field of view of WFC3, compared to previous generation instruments
such as WFPC2, NICMOS and ACS. 

In Sect.~\ref{sample} we describe the samples and the observations 
analyzed in this work, while in Sect.~\ref{mergers} we
discuss our method to classify mergers based on visual inspection of the HST images.
In Sect.~\ref{stats} we describe the statistical analysis of the results.
In Sect.~\ref{discussion} we discuss our findings and we outline our framework
for their interpretation in Sect.~\ref{implications}. Finally, in Sect.~\ref{conclusions}
we summarize our results and we draw conclusions.

Throughout the paper we assume $\Omega_M = 0.3$, $\Omega_\lambda = 0.7$ and H$_0$ = 70 Km s$^{-1}$ Mpc$^{-1}$. For the magnitude system, we use AB magnitudes.

\section{The Sample}
\label{sample}

One fundamental goal of this work is to establish whether there
is a significant difference in  the role played by mergers among 
different classes of  AGNs and for
different bins of redshift and   bolometric power. To this aim, we use well
defined samples  of classical radio  galaxies spanning 5 dex  in radio
power. We also  select new samples of both radio quiet and radio loud AGNs,
as well as non-active galaxies  at $z>1$. In doing so, we  use particular care in
separating the  radio quiet and  radio loud populations,  as explained
below.

In the  following we  describe the properties  of each of  the samples
used as part  of this work, together with  the selection criteria used
to derive new samples.

\subsection{Radio loud AGNs: $z>1$ 3CR radio galaxies}
\label{highz3cr}
The  first sample we  consider is  composed by the  radio  galaxies with
$z>1$ from  the 3CR  catalog \citep{spinrad}.  These are all powerful
radio galaxies belonging to the Fanaroff--Riley class II \citep[FR~II][]{fanaroffriley}, i.e.
those in which the brightest components of the radio structure lie at the edges of the radio source, in contrast to
the FR~Is, in which the peak of the radio emission is located at the core.

The  3CR is a  flux limited
sample selected at low radio frequencies (S$_{178} > 9$ Jy at 178MHz).
Since the radio emission at such frequencies is dominated by the radio
lobes, the selection is independent of the AGN orientation. The original catalog 
includes both Type~1 (QSOs) and Type~2 objects (radio galaxies) and it
is one of the best studied samples of radio loud AGNs, being perfectly
suitable to  test AGN unification  scenarios. The high redshift objects  included in
the  3CR are  all firmly  established  radio loud  AGNs with  powerful
relativistic jets emanating from  the central supermassive black hole.
Even at the lowest luminosities,  the active nucleus is strongly radio
loud  \citep[see  e.g.][]{sikora07,papllagn},  assuming  the  canonical
threshold  $R =  F_{5GHz}  / F_{B}  >  10$ \citep[]{kellerman89},  where
$F_{5GHz}$ and $F_{B}$  are the fluxes in the radio band  at 5 GHz and
in  the optical  B band,  respectively.  The $z>1$ sub-sample of the 3CR catalog includes 58
objects. The highest  redshift object is the  radio galaxy
3C257 at  $z=2.47$.  

The  radio power of  these AGNs is  $L_{151} \sim
10^{35}$ erg $s^{-1}$ Hz$^{-1}$  or slightly higher, which corresponds
to  a bolometric  luminosity  $L_{bol} \sim  10^{45-46}$ erg  s$^{-1}$
assuming  standard  bolometric   corrections.   This  result  is  also
supported by the  estimates of the X-ray luminosity  for some of these objects,
which  is  typically  in  the  range $L_{2-10}  \sim  10^{44-45}$  erg s$^{-1}$.
\citet[]{salvati08} observed  a sample of  the most luminous  and most
distant  radio  galaxies and  QSOs  from  the  3CR catalog  with  {\it
XMM}. They found that the  intrinsic X-ray luminosity is in the range
$6 \times 10^{44}  < L_{2-10keV} < 2\times 10^{46}$  erg s$^{-1}$, the
Type~1 QSOs  being a factor  of $\sim 6$  or higher brighter  than the
(Type~2) radio galaxies. These  authors interpreted such a discrepancy
as  a result  of the  presence  of a  beamed component  in the  Type~1
QSOs. \citet[]{wilkes13} found similar  results based on Chandra data.  
Torresi et al. (priv. comm.) analyzed both
Chandra and XMM archival data for the 3CR radio galaxies observed with
HST,  and found intrinsic  X-ray luminosities  as low  as $L_{2-10keV}
\sim  2\times 10^{44}$  erg  s$^{-1}$. However,  the absorbing  column
density is poorly  constrained, and it is possible  that at least some
of the Type~2 radio galaxies  are in fact Compton-thick. In that case,
the derived X-ray  luminosity sets a lower limit  to the intrinsic AGN
luminosity in that band. 

Twelve  3CR radio galaxies were observed
with  WFC3--IR and  the  F140W filter (in addition to WFC3-UVIS and F606W, not used in this work)
as  part  of program  SNAP13023
(PI  M. Chiaberge, Hilbert et al., in preparation). 
The wide-band filter used for the HST observations (F140W) 
includes emission lines at the redshift considered here
(mainly H$\alpha$, H$\beta$ and/or [OIII]5007, depending on the redshift of the source). 
However, the only object that is significantly contaminated by line emission is 3C~230 \citep[see][]{steinbring11}.
The HST image of this specific object only shows the narrow-line region and thus we cannot observe the stellar component of the host. 
Therefore we exclude this source from our sample\footnote{A quick inspection of the images from the HST-SNAP13023 
program shows that 3C~230 is the only object that 
presents an identical structure in both of the observed bands 
which is easily recognizable as due to emission line filaments. All other objects show, in the 
filter used in this work, a much smoother morphology that 
is not typical of emission line regions. The bandwidth of F140W is extremely wide ($\sim 0.4\mu$m), therefore the
expected emission line contamination is small. 
In fact, for all radio galaxies except for 3C~230, we estimate that, based on the observed emission line flux from 
ground based spectra, the total 
emission line contamination does not exceed ~20\% of the total observed flux (Hilbert et al., in preparation). Similar considerations
hold for the other samples used in this work.}.  
HST  snapshot  surveys of  complete
samples  are   well  suitable  for  statistical   studies,  since  the
observations  are  scheduled  by  randomly picking  objects  from  the
original target  list to fill gaps  in the HST  schedule. The observed
sample spans the  entire range of redshift of  the original list, i.e.
from  $z=1.0$   to  2.47.   The  comparison  samples described in the following are
tailored  to match  the properties  of these  3CR galaxies,  in either
bolometric power or redshift range (or both).

In Tab.~\ref{hz3ctab} we report the data for the 11 sources belonging to this sample. The 24$\mu$m luminosities
taken from \citet{podigachoski15} are used in the following to check that the relevant 
comparison sample is correctly matched to the Hz3C.

\begin{figure}
\epsscale{1.1}
\plotone{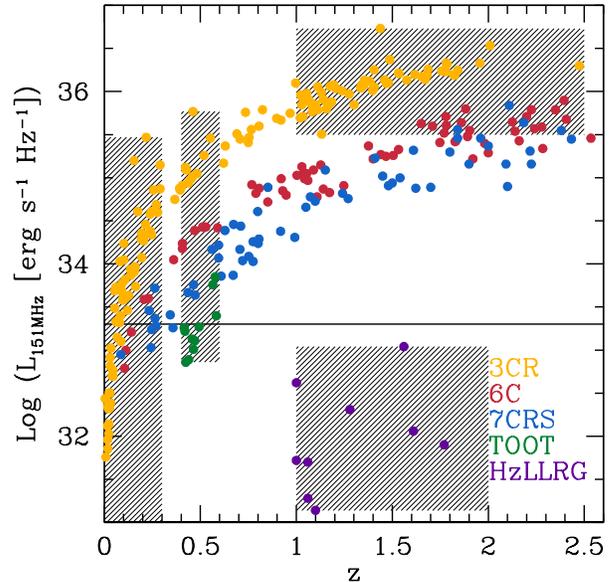}
\caption{Radio loud samples considered in this paper. The horizontal line represents the canonical separation between
low and high power radio galaxies (i.e. FR~I and FR~II, respectively). The color code identifies the different
catalogs, and the shaded area show the regions of the radio luminosity v redshift plane where HST observations 
are suitable for this paper (see text). Here, the z$< 0.3$ range is covered by the 3CR sources (yellow), and 
the four groups that are part of the Willott sample (6C, 7C and TOOT are in red, blue and green, respectively
plus the 3CR with $z\sim 0.5$) cover the range around z=0.5. Both the low-z 3CR and the Willott sample are 
described in Sect.~\ref{lowerz}. The $z>1$ range
is covered again by the 3CR at high luminosity, and by the ECDFS HzLLRGs (purple) at the lowest luminosities. 
See Sect.~\ref{sample} for more details.\label{rlsamples}}
\end{figure}

In Fig.~\ref{rlsamples} we show the location of the 3CR sample in the
radio power vs. redshift plane, with respect to other samples used
throughout the paper. Note that, at any redshift, 3CR objects are always the 
most powerful radio sources.

\subsection{Radio loud AGNs: low-luminosity radio galaxies at $1<z<2.5$}
\label{highzfr1}

The sample of high redshift low-luminosity radio galaxies (HzLLRGs) is
derived   from   the  Extended   {\it   Chandra}   Deep  Field   South
\citep[ECDFS][]{ecdfs}.   We use  the \citet{bonzini12}  catalog  and AGN
type classification to derive this sample. The criteria are as follows.

{\bf i)} The  spectroscopic, if available, or photometric  redshift of the
source must be in the range $1<z<2.5$.  The redshift range is chosen
to roughly match the range  spanned by the 3CR sample described above.

{\bf ii)} The  total radio power  at 1.4  GHz must be  greater than  $P_{1.4GHz} >
10^{30}$  erg  $s^{-1}$  Hz$^{-1}$  and below  the  fiducial  FRI/FRII
separation  $P_{1.4GHz} <  4 \times  10^{32}$ erg  $s^{-1}$ Hz$^{-1}$ \citep{fanaroffriley}.

{\bf iii)}  The  source  must  be  classified  as  a  radio  loud  AGN  by
\citet{bonzini13}, i.e.  it  must show radio emission in  excess of that
produced by starbursts.  This is measured by using
the  $q_{\rm  24obs}$ parameter,  defined  as  $q_{\rm  24obs} =  \log
(S_{\rm 24obs}  / S_{\rm  r})$, where $S_{\rm  24obs}$ is  the observed
flux density at 24$\mu$m and  $S_{\rm r}$ is the observed flux density
at 1.4GHz.  \citet{bonzini13} define an  object as radio loud if $q_{\rm
24obs}$ is 2$\sigma$ off of the starburst locus, defined using the M82
template  (see  their Fig.~2).   

{\bf iv)}  We  also  impose the  additional
constraint that its X--ray luminosity must not exceed $L_{2-10keV} = 
10^{44}$  erg s$^{-1}$.  This is  to ensure  that  the objects  are consistent  with
typical  low power  radio  galaxies  at both  low  and high  redshifts
\citep[e.g.][]{barbarafr106,tundo12}.  The  upper limit 
to the X--ray luminosity is also consistent with  the  results of \citet{terashima03}. These authors showed  that the radio
loud  -- radio  quiet  dichotomy  can be  redefined  using the  X--ray
emission instead of the flux in  the optical band. This is particularly useful for Type~2
AGNs such as the ones considered in this work, since the optical emission form the accretion disk is heavily obscured
in these object.
According to  such a scheme, powerful radio loud AGNs are present for  R$_X \gtrsim  -3.5$, where R$_X$ is defined 
as the logarithm of the ratio
between the radio luminosity ($\nu$L$_\nu$) at 1.4GHz and the X-ray luminosity L$_{2-10}$.   
At low  radio powers, it is  easy to confuse a  powerful radio quiet
AGN that  shows some radio emission  with a radio loud  object that is
intrinsically weak at all wavelengths. Note that our selection criteria 
for the radio and X-ray luminosities correspond to R$_X > -2.3$, which is
a rather conservative value. We prefer to follow a conservative approach because for low luminosity AGNs 
the transition between radio--quiet and radio--loud most likely
occurs at higher values than for high power objects, similarly to the classical 
radio--loudness parameter R derived using the radio-to-optical luminosity ratio \citep{terashima03,papllagn,sikora07}.

{\bf v)} The source must  be observed with HST
WFC3--IR.   Most of  the  images  are taken  from  the CANDELS  survey
\citep[]{koekemoer11,grogin11},   or  from  surrounding   fields  observed as part of 
program GO-12866 (see
Table~\ref{hzllrgtab}). The ECDF sample includes 6 HzLLRGs. 

We also  use the  same criteria  to select HzLLRG  in the  CANDELS UDS
field  in the  Subaru/XMM  Newton  Deep field  survey  \citep[SXDF,][]{ueda08}.   
The radio data are from  \citep[]{simpson12}, the 24$\mu$m
Spitzer IR data are taken from the Spitzer Public Legacy Survey of the
UKIDSS  Ultra Deep  Survey (SpUDS,  P.I. J.S. Dunlop) using  the  IRSA database.
Only  two galaxies  that satisfy our  selection criteria
lie in the region covered by the HST WFC3 observations.
%The photometric redshift
%of the former \citep[ID=48  in the][catalog]{simpson12} is very porrly
%contrained.   \citet{simpson12} estimates  z=1.50, with  a  very large
%uncertanty that  allows values between  z=0.99 and 2.52. On  the other
%hand,  \citet{williams08}  reports $z_{\rmphot}  =  2.2$, with  errors
%consistent with the previous values.  The value of $q_{\rm 24obs}$ for
%this  source  is very  close  to  the  region occupied  by  starbursts
%galaxies in Fig.~2 of \citet{bonzini13}.

The properties  of the 8 galaxies included  in the HzLLRG  sample are
reported in Table~\ref{hzllrgtab}. 
%{\color{red} All of the objects have R$_X > -3.5$,
%therefore they are truly radio loud, according to the Terashima et al. definition.}
In Fig.~\ref{rlsamples} we show
the radio power vs. redshift distribution of these AGNs, compared
with the other radio loud samples. Note that the two radio loud samples
at $z>1$ used in this paper (i.e. the 3CR and the HzLLRGs) are separated by
about 4dex in radio power, on average.

\begin{deluxetable*}{rcccccr}
\tablecolumns{7}
\tablewidth{0pc}
\tablecaption{High z 3CR Radio Galaxies \label{hz3ctab}}
\tablehead{
\colhead{Name} & \colhead{R.A. (2000.0)}   & \colhead{Decl. (2000.0)}    & \colhead{Redshift} & 
\colhead{$\log$ P$_{1.4\rm GHz}$}     & \colhead{L$_{24\mu m}$} \\
\colhead{} & \colhead{hh:mm:ss.ss} & \colhead{dd:mm:ss.ss} & \colhead{z}  & \colhead{$\log$[erg s$^{-1}$ Hz$^{-1}$]}  & \colhead{$\log$[erg s$^{-1}$ Hz$^{-1}$]}}\\
\startdata
3C 210   & 08:58:10.0     &         +27:50:52       &             1.169 &       35.13   & 45.65  \\
%3C 230   & 09:51:58.8     &         --00:01:27       &             1.487 &       35.66   & 45.45  \\
3C 255   & 11:19:25.2     &         --03:02:52       &             1.355 &       35.29   & $<$44.54  \\
3C 257   & 11:23:09.2     &         +05:30:19       &             2.474 &       35.94   & 45.94  \\
3C 297   & 14:17:24.0     &         --04:00:48       &             1.406 &       35.33   & 44.84  \\
3C 300.1 & 14:28:31.3     &         --01:24:08       &             1.159 &       35.37   & --  \\
3C 305.1 & 14:47:09.5     &         +76:56:22       &             1.132 &       35.10   & 45.36  \\
3C 322   & 15:35:01.2     &         +55:36:53       &             1.168 &       35.20   & 44.90  \\
3C 324   & 15:49:48.9     &         +21:25:38       &             1.206 &       35.34   & 45.48  \\
3C 326.1 & 15:56:10.1     &         +20:04:20       &             1.825 &       35.75   & 45.64  \\
3C 356   & 17:24:19.0     &         +50:57:40       &             1.079 &   34.96$^a$   & 45.52  \\
3C 454.1 & 22:50:32.9     &         +71:29:19       &             1.841 &       35.60   & 45.67  \\
\enddata
\tablecomments{S$_{1.4}$ from \citet{condon98} except for $^a$ \citet{laingpeacock80}. The 24$\mu$m luminosities are derived from the fluxes published in \citet{podigachoski15}.}
%\tablenotetext{a}{}
\end{deluxetable*}

\begin{deluxetable*}{rrrccccr}
\tablecolumns{7}
\tablewidth{0pc}
\tablecaption{High z Low Luminosity Radio Galaxies \label{hzllrgtab}}
\tablehead{
\colhead{ID} & \colhead{R.A. (2000.0)}   & \colhead{Decl. (2000.0)}    & \colhead{Redshift} & HST Survey or Prog. ID &
\colhead{$\log$ P$_{1.4\rm GHz}$}   & \colhead{$\log$ L$_{2-10}$}    & \colhead{q24} \\
\colhead{} & \colhead{hh:mm:ss.ss} & \colhead{dd:mm:ss.ss} & \colhead{z}  & \colhead{}  & \colhead{$\log$[erg s$^{-1}$ Hz$^{-1}$]} & \colhead{$\log$[erg s$^{-1}$]} & \colhead{}}\\
\startdata
\cutinhead{ECDFS}
 65 &  03:31:23.30  & -27:49:05.80     &     1.10$^a$    & 12866   &      31.85   &     $<$42.56      &        -1.54 \\
127 &  03:31:34.13  & -27:55:44.40     &     1.06        & 12866   &      30.51   &     $<$42.69      &        -0.09 \\
215 &  03:31:50.74  & -27:53:52.15     &     1.77        & 12866   &      31.13   &     $<$42.69      &        -0.46 \\
338 &  03:32:10.79  & -27:46:27.80     &     1.61$^b$    & CANDELS &      31.29   &      42.43      &        -0.98 \\
410 &  03:32:19.30  & -27:52:19.38     &     1.10$^b$    & CANDELS &      30.37   &     $<$42.27      &        -0.23 \\
412 &  03:32:19.51  & -27:52:17.69     &     1.06        & CANDELS &      30.93   &     $<$42.24      &        -0.51 \\
%693 &  03:33:06.16  & -27:48:41.65     &     1.00    &        30.95   &     $<$42.63      &        -0.93 \\
\cutinhead{UDS}                                               
  48  &  02 18 18.38 & -05 15 45.2 &  1.56       & CANDELS & 32.27  &   $<$43.93 &  0.09 \\
  124 &  02 17 04.77 & -05 15 18.1 &  1.28$^b$  & CANDELS & 31.54  &  $<$43.73 & -0.50 \\                            
\enddata
\tablecomments{For the ECDFS sources the ID corresponds to the source ID in the \citet{bonzini12} catalog. For the 
UDS galaxies, the ID is the \citet{simpson12} source number. Redshifts are photometric redshifts from \citet{bonzini12}, 
except where stated otherwise. The UDS sources are undetected in the X-rays.}
\tablenotetext{a}{Photometric redshift from \citet[]{cardamone10}}
\tablenotetext{b}{Spectroscopic redshift}
\end{deluxetable*}

\begin{deluxetable}{rrrcc}
%\tablecolumns{5}
%\tablewidth{0pc}
\tablecaption{Radio Quiet Low Power Ty2 AGN at $1<z<2.5$ \label{tablelpty2agn}}
\tablehead{
\colhead{ID} & \colhead{R.A.}   & \colhead{Decl.}    & \colhead{Redshift} & 
 \colhead{$\log$ L$_{2-10}$}    \\
\colhead{} & \colhead{hh:mm:ss.s} & \colhead{dd:mm:ss.s} & \colhead{z}   & \colhead{$\log$[erg s$^{-1}$]} }\\
\startdata
\cutinhead{Low Power Type 2 AGN (CDFS)}
125 & 03:32:06.77   & -27:49:14.10  &  1.050  & 42.04 \\
147 & 03:32:09.22   & -27:51:43.50  &  1.352  & 41.99 \\
216 & 03:32:15.26   & -27:44:38.60  &  1.109  & 41.70 \\
225 & 03:32:15.91   & -27:48:02.20  &  1.520  & 41.90 \\
226 & 03:32:16.04   & -27:48:59.90  &  1.413  & 41.91 \\
247 & 03:32:17.84   & -27:52:10.80  &  1.760  & 42.15 \\
317 & 03:32:23.16   & -27:45:55.00  &  1.224  & 42.25 \\
318 & 03:32:23.17   & -27:44:41.60  &  1.571  & 42.09 \\
321 & 03:32:23.61   & -27:46:01.40  &  1.033  & 41.98 \\
376 & 03:32:27.04   & -27:53:18.60  &  1.103  & 41.88 \\
389 & 03:32:28.62   & -27:45:57.20  &  1.626  & 41.97 \\
394 & 03:32:28.85   & -27:47:56.00  &  1.383  & 41.89 \\
416 & 03:32:29.94   & -27:52:52.80  &  1.017  & 41.76 \\
419 & 03:32:30.05   & -27:50:26.80  &  1.005  & 41.48 \\
428 & 03:32:31.11   & -27:49:40.00  &  1.508  & 41.79 \\
455 & 03:32:33.06   & -27:48:07.80  &  1.188  & 41.75 \\
462 & 03:32:33.67   & -27:47:51.20  &  1.388  & 41.84 \\
463 & 03:32:33.84   & -27:46:00.60  &  1.903  & 42.09 \\
491 & 03:32:35.80   & -27:47:35.10  &  1.223  & 42.13 \\
493 & 03:32:35.98   & -27:48:50.70  &  1.309  & 42.19 \\
504 & 03:32:36.35   & -27:49:33.40  &  1.508  & 41.81 \\
536 & 03:32:39.07   & -27:53:14.80  &  1.380  & 41.98 \\
541 & 03:32:39.42   & -27:53:12.70  &  1.381  & 42.00 \\
545 & 03:32:39.65   & -27:47:09.60  &  1.317  & 41.88 \\
558 & 03:32:41.01   & -27:51:53.40  &  1.476  & 42.19 \\
579 & 03:32:43.45   & -27:49:02.20  &  1.603  & 41.97 \\
\enddata                                               
\tablecomments{The ID corresponds to the source ID in the \citet{xue11} catalog. 
The redshifts (z$_{\rm adopt}$ in the Xue et al. catalog) are spectroscopic, if available, or photometric. 
The intrinsic X-ray luminosities (converted to the 2-10keV band as explained in the text) 
are also from \citet{xue11}.}
%\tablenotetext{a}{Photometric redshift from \citet[]{cardamone10}}
%\tablenotetext{b}{Spectroscopic redshift}
\end{deluxetable}

\begin{deluxetable*}{rrrccc}
%\tablecolumns{5}
%\tablewidth{0pc}
\tablecaption{Radio Quiet High Power Ty2 AGN at $1<z<2.5$ \label{tablehpty2agn}}
\tablehead{
\colhead{ID} & \colhead{R.A.}   & \colhead{Decl.}    & \colhead{Redshift} & 
 \colhead{$\log$ L$_{2-10}$}  & \colhead{$\log$ L$_{24\mu m}$}   \\
\colhead{} & \colhead{hh.mm.ss.s} & \colhead{dd.mm.ss.s} & \colhead{z}   & \colhead{$\log$[erg s$^{-1}$]} & \colhead{$\log$[erg s$^{-1}$]} }\\
\startdata
%\cutinhead{High Power Type 2 AGN (CDFS)}
166 & 03:32:10.93  & -27:44:15.20 & 1.605 & 44.21  & 44.81 \\
243 & 03:32:17.19  & -27:52:21.00 & 1.097 & 44.05  & 44.39 \\
257 & 03:32:18.35  & -27:50:55.61 & 1.536 & 44.07  & 44.30 \\
278 & 03:32:20.07  & -27:44:47.51 & 1.897 & 44.06  & 45.04 \\
351 & 03:32:25.70  & -27:43:06.00 & 2.291 & 44.31  & 45.46 \\
518 & 03:32:37.77  & -27:52:12.61 & 1.603 & 44.23  & 45.38 \\
577 & 03:32:43.24  & -27:49:14.50 & 1.920 & 44.12  & 44.70 \\
713 & 03:33:05.90  & -27:46:50.70 & 2.202 & 44.02  & --    \\
720 & 03:33:07.64  & -27:51:27.30 & 1.609 & 44.39  & --    \\
\enddata                                               
\tablecomments{The ID corresponds to the source ID in the \citet{xue11} catalog. 
The redshifts (z$_{\rm adopt}$ in the Xue et al. catalog) are spectroscopic, if available, or photometric. 
The intrinsic X-ray luminosities (converted to the 2-10keV band as explained in the text) 
are also from \citet{xue11}. The 24$\mu$m luminosities are from \citet{cardamone08}.}
%\tablenotetext{a}{Photometric redshift from \citet[]{cardamone10}}
%\tablenotetext{b}{Spectroscopic redshift}
\end{deluxetable*}

\subsection{Radio quiet AGNs: low-power Type~2 AGNs at $1<z<2.5$}
\label{lpqso2}
The sample  of low-power  Type~2 AGN (LPTy2AGN)  is selected  from the
4Msec CDFS catalog and  source classification \citep[]{xue11}.  We
include objects that are classified as AGNs, that do not show strong broad emission lines in the optical spectrum, and 
whose redshift is $1.0 <z<2.5$.  We also use the additional constraint that the intrinsic
(de-absorbed) X-ray  luminosity (integrated between  2 and 10  keV, in
the rest frame of the  source) must be $L_{2-10keV} < 2\times 10^{42}$
erg s$^{-1}$,  in order  to be consistent  with the properties  of the
corresponding  radio loud  sample  described in  Sect.~\ref{highzfr1}. The L$_{2-10keV}$ luminosity
is derived from the 0.5-8keV luminosity listed in the catalog, converted to the 2-10keV band
using a photon index $\Gamma = 1.8$.
Furthermore, for the objects that are detected in the radio, we require that
the X-ray radio loudness parameter $R_X$ defined in Sect.~\ref{highzfr1} is $<-3.5$.
The sample includes 26 objects. 
These are  very low power  AGNs that are
similar to  Seyfert~2 galaxies in the local  Universe.  The properties
of  these AGNs  are reported  in Tables~\ref{tablelpty2agn}.  The HST/WFC3-IR
images for this sample are taken from the CANDELS survey \citep[]{koekemoer11,grogin11}.

\subsection{Radio quiet AGNs: high-power Type~2 AGNs at $1<z<2.5$}

As for the low power AGNs described above, the sample of high-power Type~2
AGN (HPTy2AGN) is drawn from the CDFS.  When selecting this sample we
want to  match as close  as possible both  the redshift range  and the
bolometric  luminosity  of  the  powerful  3CR  radio  galaxy  sample.

The sample of high-power
Type~2 AGNs includes objects  classified as AGN with redshift
$1<z<2.5$  and  intrinsic  X-ray  luminosity L$_{2-10keV}  > 10^{44}$ erg  s$^{-1}$. 
The lower limit  in X-ray power  is chosen to
match   the    properties   of    the   3CR   sample    described   in
Sect.~\ref{highz3cr}. Two of the objects (166 and 577) show relatively broad MgII and CIII] emission lines in
the optical spectrum \citep[]{szokoly04}. However, the HST images clearly show the host galaxy and there is no evidence for the presence of any strong
unresolved nuclear source, as it would be expected for a powerful Type~1 QSO. 
Therefore, for the purpose of this work, we consider those two objects as Type~2,
regardless of the presence of broad lines.

As for the LPTy2AGN, for the objects that are detected in the radio, we require that
the X-ray radio loudness parameter $R_X$ defined in Sect.~\ref{highzfr1} is $<-3.5$.
This   sample    includes    9   AGNs    (see
Table~\ref{tablehpty2agn}).  The HST/WFC3-IR images  for this sample are taken
from the CANDELS survey \citep[]{koekemoer11,grogin11}.

Since the X-ray luminosity might be poorly constrained because of the uncertainty on the 
amount of nuclear obscuration, 
we should also make sure that this sample and the Hz3Cs have similar mid-IR luminosities. We performed 
a K-S test to check whether the distributions of 24$\mu$m luminosity of the Hz3C and HPTy2AGN 
are different\footnote{Note that 
we did not perform the same test for the LPTy2AGN because those are all undetected at 24$\mu$m.}. 
The resulting p-value is p=0.1, therefore we cannot reject the null hypothesis that the two distributions are drawn from
the same population.

\subsection{Non-active galaxies at $1< z < 2.5$}
\label{nonactive}
This sample is derived from the 3D--HST survey of the GOODS--SOUTH field
\citep[]{giavalisco04,brammer12,skelton14}. We select galaxies with spectroscopic redshift
$1< z < 2.5$ and with magnitude  in the F140W filter $19 < m_{F140W} <
22$. The magnitude range is chosen  to match the magnitudes of the 3CR
radio galaxies from Sect.~\ref{highz3cr} (Hilbert et al., in prep.).  The full
sample includes  145 galaxies. We limit  the sample to  the 50 objects
included in the southern half area of the field  observed with WFC3-IR
and the F140W  filter as part of the 3D-HST  survey. 
Five of these objects are AGNs, therefore we are left with a sample of 45 galaxies.
These are sufficient to provide us
a statistically  sound sample  to  be compared  with the  active
galaxy samples  described above. We refer to this sample as the {\it bright} sample of non-active galaxies.
Note that we could in principle derive a larger sample by using the full area covered by the 
3D--HST image of the GOODS-S field.
However, this would not significantly improve the statistics for a sample 
that is already the largest we consider.

In the same area of the sky, we also select a comparable sample of fainter galaxies 
($22 < m_{K_s} < 24$) from the same HST survey. This sample spans the same magnitude range as 
the bulk of our radio--quiet AGN population. In the following, we will refer to this sample
as the {\it faint} galaxies sample.

Even if we are selecting objects in the very same redshift range, the near-IR (rest-frame optical) magnitude 
might not be a good tracer of the stellar mass because of the possible presence of obscuration in some objects. 
We do not believe that this might significantly affect our results. However,
in order to perform a sanity check, we also select
a sample of non-active galaxies in the full GOODS-S area of 3D--HST matched to the stellar
mass estimates of the Hz3C galaxies. At $1<z<2.5$ these are typically between $1\times 10^{11}$ and $5\times 10^{11}$ 
M$_\sun$ \citep{seymour07}, 
although a small number of lower mass objects are also present. Although stellar mass measurements based on SED fitting
heavily rely on models, it is important to check that our non-active galaxy samples are correctly representing 
the population of objects we need for comparison.
If we restrict our range of masses between the above values, 
and after excluding those that are AGNs, we obtain 23 objects. About half of them are in common with the {\it bright} sample.
While we do not consider this as our main control sample of non-active galaxies,
results for this sample are briefly discussed in Sect.~\ref{stats}, for the sake of completeness.

The HST/WFC3-IR images  are taken as  part of the 3DHST survey with the F140W filter.

\begin{figure*}
\epsscale{1.0}
\plotone{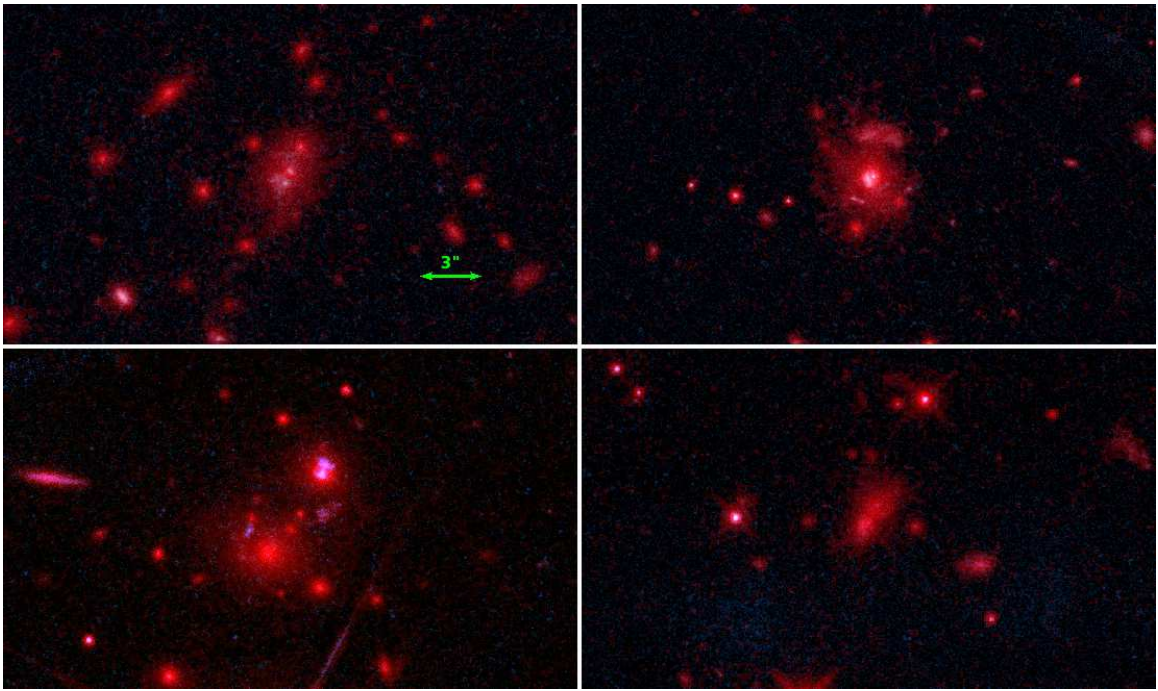}
\caption{RGB images of four radio galaxies from the high-z 3CR sample showing clear evidence for recent or ongoing major 
merger. The objects are (from left to right, top to bottom) 3C~210, 3C297, 3C356 and 3C454.1. The HST WFC3-IR F140W image was used for the R channel. The WFC3-UVIS F606W image was used for both the G and B channels. North is up, East is left. Data from Hilbert et al. (in preparation) \label{3crmergers}}
\end{figure*}

\begin{figure}
\epsscale{1.0}
\plotone{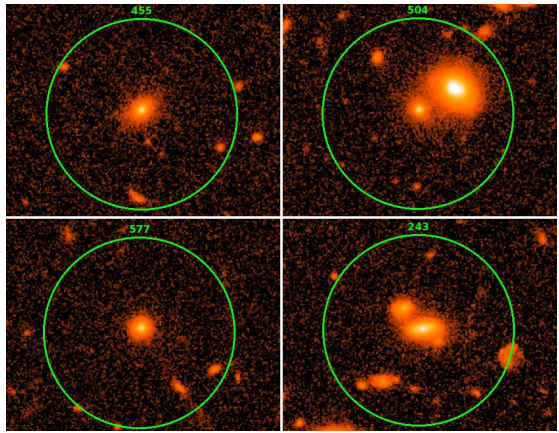}
\caption{Four examples of the morphologies observed among the $z>1$ radio quiet AGNs. In the top panels, we show two
LPTy2AGNs. A non-merger and a merger are shown on the left and on the right, respectively. 
In the bottom panels, the same is shown for two HPTy2AGNs. The circles are 6 arcsec radius, 
which correspond to $\sim 50$ kpc at the redshift of the objects. Images are from 
CANDELS \citep[HST/WFC3/F160W]{koekemoer11,grogin11}.\label{4rqagn}}
\end{figure}

\begin{figure*}
\epsscale{1.0}
\plotone{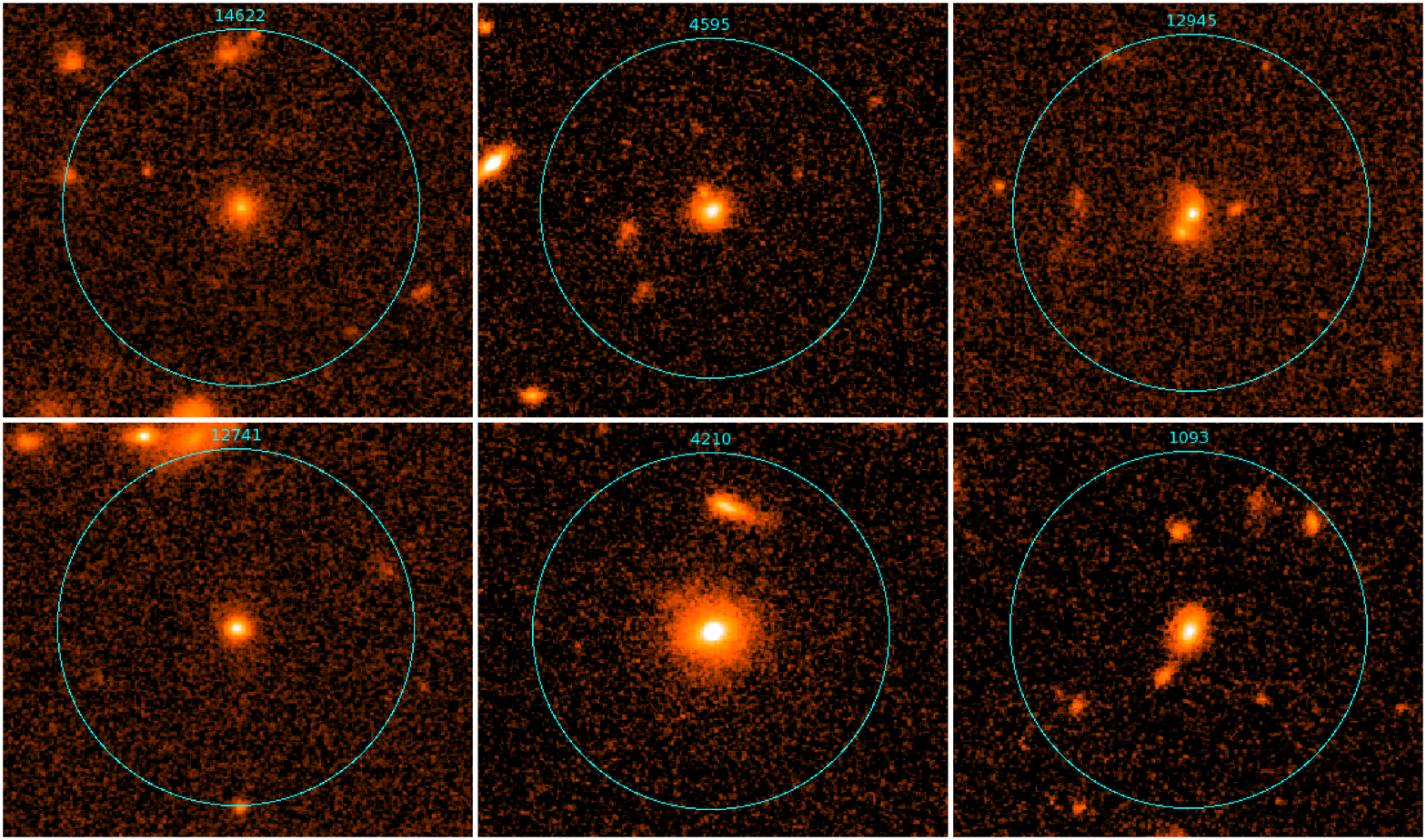}
\caption{Six examples of non--active galaxies from our $z>1$ sample. Both in the top and bottom rows, 
objects that are unanimously classified as non-mergers or
mergers are shown in the left and right panels, respectively. Objects with mixed classification are 
shown in the middle panels. The circles are 6 arcsec radius, 
which correspond to $\sim 50$ kpc at the redshift of the objects. Images are from 3D-HST  
\citep[HST/WFC3/F140W]{giavalisco04,brammer12,skelton14}\label{6nonactive}}
\end{figure*}

\section{Merger fraction}
\label{mergers}
In order  to measure  the merger  fraction we  use at least four
human classifiers for each sample.  For the Hz3C and the Ty2 AGN samples we used six classifiers, 
to be sure that the  authors of the paper were not biased {\it a priori} towards any specific result.
Each classifier visually inspects all of
the  targets.
%focusing  on  a circular area with a $\sim  50$kpc 
%projected  radius,   at  the  redshift  of  the   source.  
While  being
qualitative in nature, visual classification has been proven to be an
effective way of  classify mergers, since the eye can  pick all of the
different signatures of such  events.  On the other hand, quantitative
methods using the  Gini coefficient G, the concentration  index C, the
asymmetry index  A, or the  second-order moment of the  brightest 20
per cent of  the light (M$_{20}$) only select  particular mergers that
each of the above indicators are able to identify \citep[]{lotz11}. Therefore, if we use
any of those methods we would derive smaller merger fractions.

We  classify  objects  as   mergers  (or  post-mergers)  if   clear
signatures  of a  mergers are  present. These include the presence of double/multiple nuclei,
close pairs, 
tidal  tails, bridges, or distorted morphologies clearly indicating a recent or ongoing merger. Close pairs are defined using a 
projected separation of less than 25kpc between the center of the galaxies involved, 
corresponding to 3 arcsec in the redshift range spanned by our $z>1$ 
samples\footnote{Between z=1 and z=2.5 the projected scale varies by less than 0.1 arcsec.}. Such a scale is similar to the typical
separations observed in galaxy pairs at low redshifts \citep[e.g.][]{smithheckman89,behroozi15}\footnote{Note that the definitions of ``double nucleus'' and ``close pair'' are only formally different. 
For the purposes of our work, these objects are in fact the same. 
The different names only identify the appearance of the object in the image, since our classification is based on visual inspection. 
We classify the object as a double (or multiple) nucleus if the two objects are not clearly separated and they appear to lie within 
a common envelope. 
If the galaxies are well separated, but their nuclei are less than 
$\sim 25$kpc apart, then we call it a close pair. However, it is important to note that 
all of these objects are required to display evidence of bridges or asymmetries to be classified as mergers.}.
%{\bf Some of the objects 
%were originally classified as mergers only based on the presence of multiple companions within an area of  $\sim 50$kpc radius around the target. 
%We re-inspected all of the images and 
%removed those objects from the final counts, leaving only those that showed bridges between group members, or other signatures of
%a merger, as discussed above.}

When  possible,
mergers are  also visually distinguished between {\it  major} and {\it
minor}, assuming the usual  separation at a $\sim 4$:$1$ mass/apparent
brightness  ratio  \citep[e.g.][]{lotz11}.    If  there  is  no  clear
evidence for any of the above  properties, we classify the object as a
{\it non-merger}.   For the AGN samples, a  {\it blind} classification
is performed, i.e.  classifiers did not know whether  each object was
radio-loud or radio-quiet, and if it belonged to the high or low-power
class. Classifiers are also asked to classify objects more than once (typically twice).

While substantial agreement exists between the different classifiers for most of the objects, different 
people may  see different features in each image,
thus  the  classification for  single  sources  may  differ. This is why we do not report the merger classification for
each galaxy in  the tables.  In order to  reduce the effects of large deviations among
the different classifiers we calculate  10\%-trimmed means for
each sample\footnote{Trimmed (or truncated) means are robust estimators of central tendency, 
and it is less sensitive on the  outliers than the mean. Trimmed means are 
derived by calculating the mean after discarding parts at the high and low ends of a probability distribution. In our case, 
the distribution is defined by the number of mergers obtained for each sample by each of the classifiers, and we reject 10\% at 
both ends. Note that a 50\% trimmed mean would correspond to the median.}. 
We report the results for  each of the samples 
at $z>1$ in Tab. \ref{tabmergers}.

Examples of the morphologies observed in the different samples are given in Figs.~\ref{3crmergers}, \ref{4rqagn}, and \ref{6nonactive}.
In order to avoid confusion between our merger classification criteria, we use the figures to give a few specific examples. 
3C~297 and 3C356 in Fig.~\ref{3crmergers} 
are both classified as mergers based on those clear signatures of interaction. 
3C~297 also shows a double nucleus. Although not all cases are as straightforward as these, 
similar considerations can be made for objects in the other samples (see e.g. the systems shown in the right panes of 
both Figs.~\ref{4rqagn} and \ref{6nonactive}). In particular, the objects shown in Fig.~\ref{4rqagn} 
are most likely examples of dry mergers in which the galaxy isophotes are highly asymmetric \citep[see e.g.][]{bell06}.
The central panels of Fig.~\ref{6nonactive} show objects that some of the classifiers classified as mergers, while others did not. 
Those that classified the objects as mergers noticed some asymmetries in the isophotes of the galaxies, in addition to the  presence of
small companions.

The observed merger fraction in the radio-loud samples (Hz3C and HzLLRG)
are clearly larger than those found for radio-quiet and non-active galaxies.
In Sect.~\ref{stats}, we test this result using careful statistical analysis.

\begin{deluxetable}{lcc}
\tablecolumns{3}
\tablewidth{0pc}
\tablecaption{Observed Merger Fractions (trimmed means) for the $z>1$ samples \label{tabmergers}}
\tablehead{
\colhead{Sample} & \colhead{Sample Size}   & \colhead{Merger Fraction}} \\
\startdata
Hz3C &   11 &   100\%    \\
HzLLRG &  8 &    88\% \\
LPTy2AGN & 26   &  38\% \\
HPTy2AGN & 9 &  33\% \\
Bright Galaxies & 45 & 27\%  \\
Faint Galaxies & 46 & 20\% \\
RL (3C + LLRG)      &  19 & 95\% \\
RQ (HP+LPTy2AGN)      &  35 & 37\% \\

\enddata
%\tablecomments{}
%\tablenotetext{a}{}
\end{deluxetable}

\subsection{On the impact of different sensitivities on the merger classification}

The goal of this work is to identify merging systems in samples of objects that were observed as part of 
different surveys or observing programs. All of the images were taken using the WFC3-IR camera. Its 
extremely high sensitivity and spatial resolution allows us to detect low surface brightness features that characterize recent
merger events at z $\sim 1-2$ even in short (1 orbit or less) observations. However, for the sake of clarity, we
report in Table~\ref{surfbrightlim} the 5$\sigma$ surface brightness limits for the different surveys we use.
To derive the limits we used the WFC3-IR Exposure Time Calculator (ETC). We assume a 2x2 pixel extraction area and a spectrum
of an elliptical galaxy with a UV upturn to perform the conversion between F160W (used for the CANDELS observations) and 
F140W magnitudes. Although the CANDELS deep images we used for the radio quiet samples and for the HzLLRGs are deeper 
than all other data, we checked that the merger classification is not different if the shallower 3D-HST images are considered. 
The images show the very same features, irrespective of the redshift of the source. 
The faintest tidal structures are less prominent in the shallower images, 
but that does not significantly affect our classification. The short exposure times of the 3C snapshot survey are also sufficient 
to detect the faint features we are interested in, in the range of redshift of our sample, as shown by the fact that
that sample has the greater observed merger fraction.

\begin{deluxetable*}{lcccc}
\tablecolumns{5}
\tablewidth{0pc}
\tablecaption{Surface brightness limits \label{surfbrightlim}}
\tablehead{
\colhead{Survey or HST Prog. ID} & \colhead{Sample} & \colhead{HST Camera/Filter}   & \colhead{5$\sigma$ Surface Brightness Limit} & \colhead{Sensitivity} \\
\colhead{} & \colhead{}& \colhead{} & \colhead{$\mu_{\rm{F140W}}$ [ABmag arcsec$^{-2}$]} & \colhead{AB mag}} \\
\startdata
SNAP-13023   & Hz3C           &  WFC3/F140W    &    23.9     &  25.8       \\
CANDELS wide & RQ and HzLLRG  &  WFC3/F160W    &    24.1     &  26.5       \\
CANDELS deep & RQ and HzLLRG  &  WFC3/F160W    &    25.6     &  27.2       \\
3D-HST       & non-active     &  WFC3/F140W    &    24.1     &  26.2       \\
GO-12866     & HzLLRG         &  WFC3/F160W    &    25.0     &  26.7       \\
\cutinhead{Low-z Samples}  
GO-9045      & 3CRR,6C,7C,TOOT & WFPC2/F785LP   &   20.7     &  24.9       \\
SNAP-10173     & 3CR (z$<0.3$)     & NICMOS/NIC2/F160W & 22.5   &  24.3      \\
\enddata
\tablecomments{The 5$\sigma$ limits are estimated using the WFC3-IR ETC and assuming a 2x2 pixel extraction region. 
In column 4 we report the limit surface brightness for each datasets, converted to the WFC3-IR F140W filter to allow for an easy comparison between the different surveys. An elliptical galaxy spectrum with UV upturn 
redshifted to the appropriate redshift for each sample  was used for the conversion between the two WFC3 filters. 
RQ in the samples column refers to both the LP and HPTy2AGN samples. For the low redshift samples (GO9045 and SNAP10173)
we used {\it synphot} to convert magnitudes to the WFC3 filter system. 
In column 5 we report the 5$\sigma$ image sensitivities calculated for point sources in each of the bands used for the observations.}
%\tablenotetext{a}{}
\end{deluxetable*}

\section{Statistical Analysis}
\label{stats}
The main goal of this work is to investigate if mergers are associated
with AGN activity and if that plays a role in triggering such
a  phenomenon.  Most importantly,  we want  to test  whether different
classes  of AGNs  (e.g.   radio quiet  vs.   radio loud,  low vs  high
bolometric luminosity) are more likely  to be triggered by merger than
others.  We also  test the hypothesis that AGNs  are no different than
non-active galaxies.

We perform a set of statistical tests to compare the 
derived merger fractions for the five different samples described in
Sect.~\ref{sample}.  Throughout the paper, statistical   tests  are  performed using  
different techniques in the R environment \citep{rprogram}.

We use the Bayesian version of the proportion test using {\it bayes.prop.test}, as part of
the {\it Bayesian First Aid} package for $R$ \citep[]{bayesfirstaid}. In principle,
Bayesian tests are more useful than classical proportion tests, since they provide an estimate of the 
relative frequency of success \citep[e.g.][]{bolstad07}. In this simplified version of the Bayesian tests, the priors are 
uninformative, i.e. a uniform distribution. This is suitable for our purposes, since we have no
{\it a priori} knowledge of the distribution of mergers in each sample\footnote{We also perform a complete
set of classical proportion tests using the R task {\it prop.test}. For these classical tests we reject the null hypothesis that the
merger fractions in two samples are the same if the {\it p-value} is $p<0.01$.
Not surprisingly, the results are perfectly in agreement with the Bayesian relative frequencies. 
For the sake of clarity, here we only discuss in details the Bayesian results.}.

In table \ref{bayes} we list the relative frequencies
of success for each sample, together with the 95\% credible intervals.

By testing each sample against each of the others, 
we can firmly establish  that the
Hz3C has a larger merger fraction than both of the Ty2AGN samples (P$>99.9\%$). 
The Hz3C merger fraction is also larger than that of  the 
non-active  galaxies (both bright and faint samples, P$>99.9\%$). 
Furthermore, the HzLLRGs are significantly more associated with mergers than the 
non-active galaxies (P$>99.9$).
All other tests are inconclusive\footnote{For a 2-sample proportion test in 
classical statistics this means that
we cannot reject the null hypothesis that the observed merger fractions 
in two samples are the same ({\it p-value} $< 0.01$). 
For our Bayesian treatment, this implies that
the probability that one sample has a higher (or lower) 
merger fraction than the other is less than 99\%.}.
However, these are  very important results.

Furthermore, if we merge the two samples of $z>1$ radio loud objects and we test them against
the  radio quiet  AGNs,  the  result is  extremely  robust. The  merger
fraction in our sample of radio--loud AGNs is significantly
higher than that in the radio--quiet sample. This is again a notable result, since the  (Bayesian) 
merger fractions for the
RL and  RQ sample are 92\%  and 38\%, respectively.   As expected, the
same result  holds (with an even higher  statistical significance) for
the RL  sample against  the non--active (both bright and faint) galaxies, while the  RQ sample
does not show any statistically  significant difference with respect to
the non-active galaxies. It is particularly important to note the results for the samples that are matched in magnitude. 
On the one hand, the RL and the {\it bright} galaxy sample
are different, and on the other hand, the RQ and the {\it faint}
galaxy sample are statistically indistinguishable.

As a sanity check, to avoid possible biases from a selection made using near-IR magnitudes, 
we also used  the smaller sample of {\it high-mass} non-active galaxies (see Sect.~\ref{nonactive}). 
In particular, given that we measure a merger fraction of 30\%$^{+19\%}_{-17\%}$, the  
{\it high-mass} sample is
statistically different from the Hz3C, with a Bayesian probability $> 99.9\%$. On the other hand, 
the merger fraction in such a sample is
statistically indistinguishable from those in 
both the {\it faint} and the {\it bright} samples of non-active galaxies.

%We also checked that the Bayesian probability that each group has a higher (or lower) 
%merger fraction than that of other samples is in agreement with
%the {\it p-values} we obtain with classical tests of proportions. 

Finally, we wish to point out that among the samples of radio-loud AGNs at $z>1$ the
large majority of the objects ($\sim 90\%$) appear to be associated with major mergers. On the
other hand, for radio quiet AGNs at $z>1$, only $\sim 50\%$ of the observed mergers are major mergers. 
Note that this is only based on our qualitative visual classification of the mergers.
A more careful classification would require 2-D galaxy modeling and SEDs to derive stellar mass estimates for the
galaxies involved in each merger, which
is beyond the goals of this work.

\begin{deluxetable}{lcc}
\tablecolumns{3}
\tablewidth{0pc}
\tablecaption{Estimated Bayesian probabilities for the $z>1$ samples.\label{bayes}
}
\tablehead{
\colhead{Sample} & \colhead{Merger Fraction} & \colhead{95\% Credible Interval}} \\
\startdata                   
Hz3C             &  94\%  &  0.78  -   1.00      \\
HzLLRG           &  82\%  &  0.57  -   0.99      \\
LPTy2AGN         &  39\%  &  0.22  -   0.57      \\
HPTy2AGN         &  36\%  &  0.11  -   0.64      \\
Bright Galaxies  &  27\%  &  0.15  -   0.40      \\
Faint Galaxies   &  20\%  &  0.10  -   0.33      \\
%High-mass Galaxies & 30\% &  0.13  -   0.49      \\
RL (3C + LLRG)   &  92\%  &  0.78  -   1.00       \\
RQ (HP+LPTy2AGN) &  38\%  &  0.23  -   0.54       \\
\enddata
%\tablecomments{}
%\tablenotetext{a}{}
\end{deluxetable}

\begin{deluxetable}{lrcc}
\tablecolumns{4}
\tablewidth{0pc}
\tablecaption{Estimated Bayesian probabilities for the $z<1$  radio loud AGN samples. \label{willott}}
\tablehead{\colhead{Sample} & \colhead{Size}& \colhead{Merger Fraction} & \colhead{95\% Credible Interval}} \\
\startdata                   
\cutinhead{Willott Samples}
3CRR      &  13     &  74\%  &  0.53  -   0.94      \\
6CE       &  7      &  68\%  &  0.38  -   0.93      \\
7CRS      &  9      &  65\%  &  0.37  -   0.89      \\
TOOT      &  12     &  65\%  &  0.39  -   0.87      \\
Full Sample & 41    &  70\% &  0.56 - 0.83 \\
\cutinhead{Low Redshift Radio galaxies}
$z<0.3$ 3CR &  101 & 88\% & 0.81 - 0.94 \\      
\enddata
%\tablecomments{}
%\tablenotetext{a}{}
\end{deluxetable}

\subsection{Lower redshift samples of radio loud AGNs}
\label{lowerz}
One important  goal of  this work is  to establish whether  the merger
fraction for radio-loud AGN  depends on either redshift or luminosity.
In  order to  test  if that  is  the case,  we  used additional  lower
redshift  samples, for comparison with the $z>1$ data.    

\citet[]{mclure04}  observed  a   sample  of  41
intermediate (z$\sim 0.5$) radio galaxies with HST and WFPC2 (GO 9045,
P.I.  Willott).  The sample  is  taken  from  four different  complete
catalogs, spanning about  4 dex in radio power.  The four catalogs are
all complete, flux limited and radio selected at low frequencies.  The
sample   includes   objects   from   the  3CRR,   6CE,  7CRS and  the  TexOx-1000 (TOOT  sample,
hereinafter) \citep[][and references therein]{mclure04}. The redshift range spanned by the
objects observed with WFPC2 is between $z=0.40$ and 0.59. We will refer to these galaxies as 
the Willott sample throughout the paper.

For these objects, we retrieved 
the HST data from the MAST and we reduced them using {\it Astrodrizzle} \citep[]{astrodrizzle}. We
then classify the objects between mergers and non-mergers,
since no classification was provided in the original paper by \citet[]{mclure04}, and for
consistency with the other samples. 
Note that the filter used for these observations (F785LP) provides images
at $\sim 5300$\AA~~in the rest frame. This is similar to the rest frame wavelength
of the WFC3-IR observations for the $z>1$ samples described above. In Tab.~\ref{willott}
we report the merger fractions for this sample. The merger fractions are statistically compatible with
those for the radio-loud samples at higher redshifts.

At even lower redshifts, we use the observations of the 3CR sample with $z<0.3$ taken 
with HST-NICMOS (program SNAP10173) and the F160W filter \citep[]{madrid06}. Although these observations 
are in the rest-frame IR, they are better suitable for our purposes than the optical
data of the same sample taken with WFPC2 \citep{dekoff96,martel99} because the NICMOS images are significantly deeper.
For this sample, we consider the results published by \citet[]{floyd08}.
These authors classified all of the objects in mergers, pre-mergers, tidal-tails, 
major and minor companions. The field of view covered by NIC2 is 19.2\arcsec $\times$ 19.2\arcsec, 
which corresponds to ~50x50 kpc$^2$ at the median redshift of the sample. 
They found that 89 out of 101 objects fall into at least 
one of these categories. This corresponds to a merger fraction of 0.88, with a Bayesian 
95\% credible interval [0.81--0.94]. If we exclude objects with only {\it minor} companions (i.e.
$\gtrsim 1$ mag fainter than the radio galaxy host, as defined by Floyd at al., 2008), the fraction is reduced to 0.76 with a Bayesian 
95\% credible interval [0.68--0.84].

Since the observations are not homogeneous, these two lower redshift samples 
are only considered here for comparison. However, we do not base our conclusions 
on these samples only.

\section{Discussion}
\label{discussion}
The main result of this work is that the samples of radio quiet and radio loud Type~2 AGNs 
at $1<z<2.5$ have different merger fractions. 
We showed that there is clear statistical evidence that the radio loud AGNs almost
always reside in environments where mergers are undergoing, or that recently happened. 
As discussed below, this is independent of the radio (or bolometric) power of the AGN, and it is statistically
compatible with merger fractions as high as $\sim 100$\%. 

The sample with the highest observed merger fraction is the $z>1$ 3CR. In principle, this sample
could be biased, since  HST only imaged $\sim 35$\% of the radio galaxies with $z>1$ included in the 3CR catalog.
However, it is important to note that the observations were performed 
as part of an HST SNAPSHOT program. In SNAP programs, targets are randomly selected based
on the availability of gaps in the HST schedule. Therefore, there was no {\it a priori} knowledge of the 
properties of the observed targets with respect to the complete sample. 
But since the observed sample is small, it is still possible that we ended up picking a
large number of objects in mergers only by chance. In the following we briefly discuss such an issue.
For example, we could in principle assume that the original population is
composed by $\sim$50\% of mergers and 50\% of non-mergers, similarly to what is 
observed in our sample of non-active galaxies. We can test the probability of obtaining
a sample of 12 mergers, randomly extracted from the complete sample of 34 radio galaxies 
included in the 3CR catalog with $z>1$. 
The selection of the 3CR catalog only covered part of the sky (Decl. $> -10$ deg). Therefore, we should 
correct for the area coverage to obtain the total number of radio sources (i.e. 58 radio galaxies). 
The binomial probability of obtaining 12 mergers out of 12 observed
is extremely small ($P=2\times 10^{-4}$). 
However, when objects are randomly drawn
from a sample without replacement, the Hypergeometric distribution must be used instead of the binomial. 
In that case, the probability of obtaining 12 mergers out of 12 observed, from 
a population of 58 objects in which 50\% of the objects are mergers is $P=5.8\times 10^{-5}$. 
Note that if the merger fraction in the 
complete sample is higher than 72\%, then the (hypergeometric) probability of obtaining 12 mergers
is $P > 0.01$. Since that corresponds to the significance level we set for all of the statistical tests,
we can state that we cannot completely rule out that the merger fraction
in the 3CR is as low as $\sim 72\%$. Interestingly, this is similar to the lower 
value of the credible interval given by the Bayesian analysis for this sample (see Tab.~\ref{bayes}).
The only other selection bias that might affect our sample is the \citet[]{eddington913} bias. However, that
would have the effect of lowering the actual number of sources in our sample, and that in turn would lower the 
probability of the hypergeometric distribution.

Therefore, we conclude that even if the number of 3CR galaxies observed at $z>1$ is small,
the probability that the observed merger fraction is overestimated because of any selection bias is extremely
small. 
%And even if the true merger fraction in the all-sky sample of powerful radio galaxies was as low as $\sim72\%$, this would not
%affect our conclusions.

\begin{figure*}
\epsscale{1.1}
\plottwo{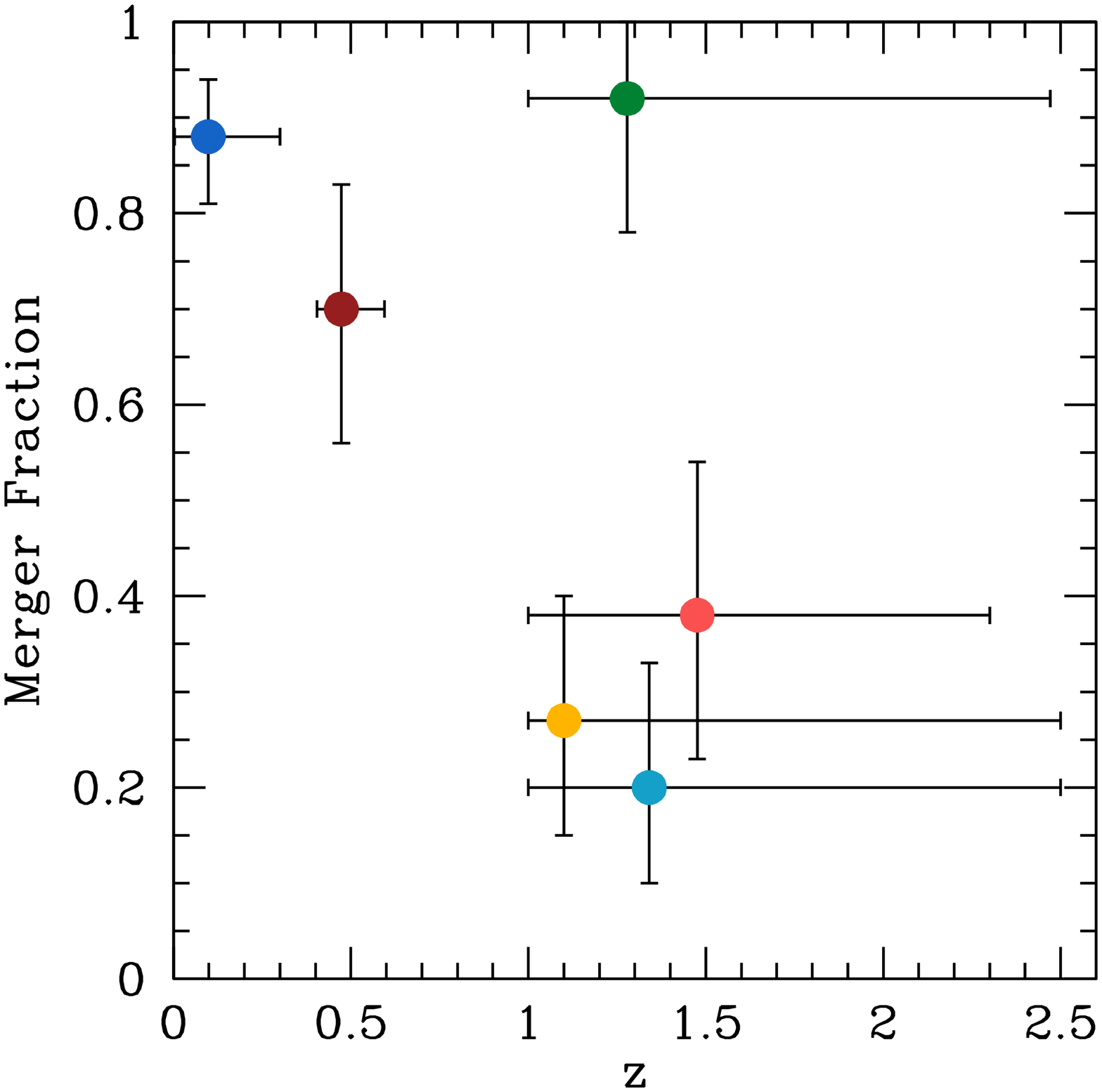}{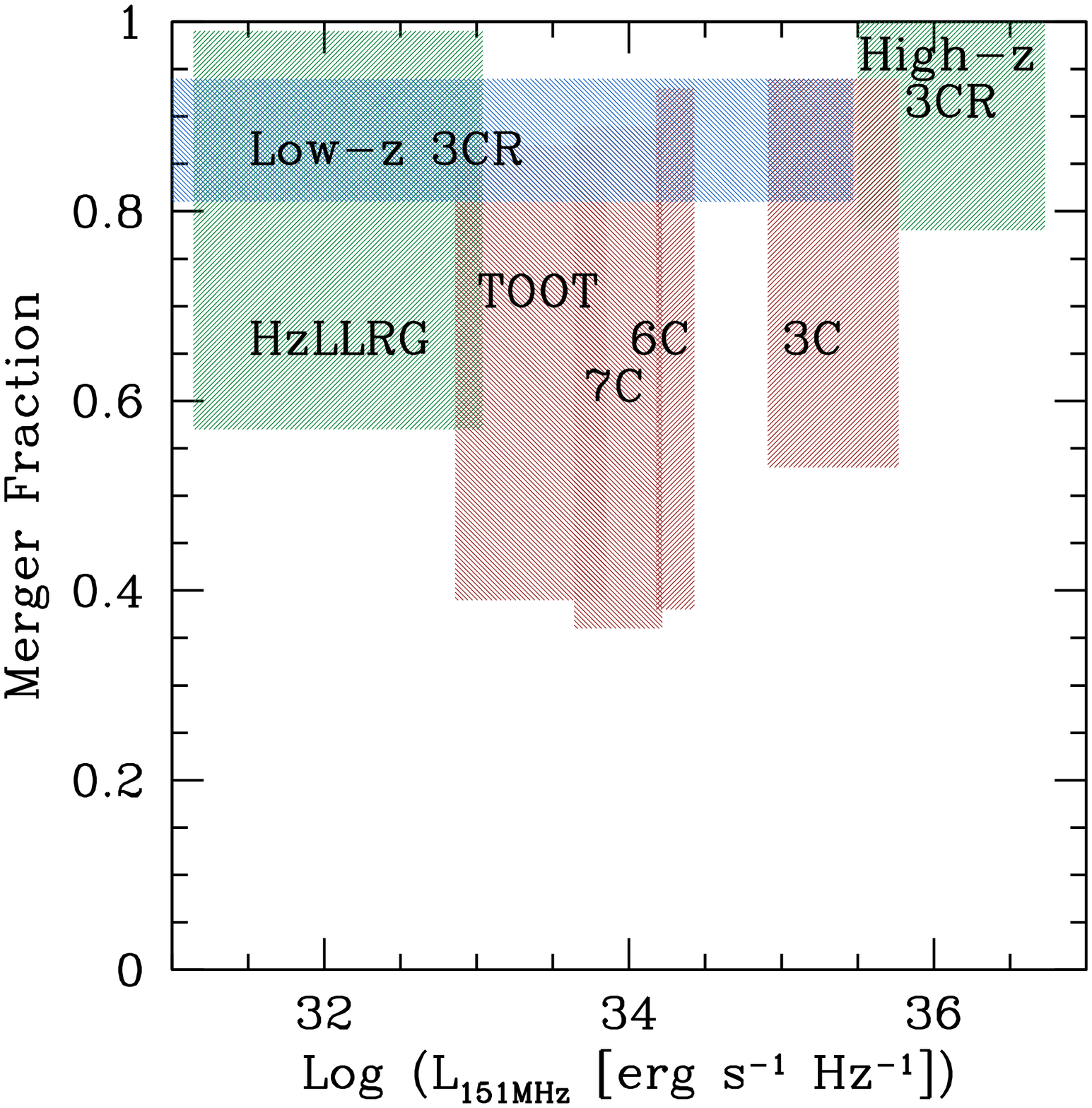}
\caption{{\bf Left:} merger fraction vs. redshift for the samples of radio loud and radio quiet AGN, 
and for non-active galaxies. The coordinates of the filled dots are the median redshift of each sample, 
and the Bayesian merger probability. The errorbars show the range of redshift spanned by each sample and the Bayesian 
95\% credible intervals for the merger probability. The High-z radio--loud sample (green dot) corresponds to the 
3CR at $z>1$  merged with the HzLLRG sample to improve the statistics. The radio-quiet Ty2 AGN samples at $z>1$ are also merged 
(light red). 
The bright and faint samples of non active galaxies are plotted in yellow and cyan. 
The Willott sample (3CR, 6CE, 7CRS and TOOT samples at $z\sim 0.5$) is in red. 
The blue dot represents the 3CR sample with $z<0.3$. 
{\bf Right:} merger fraction vs. radio luminosity at 151MHz for the radio loud samples. 
In this panel, the Willott samples (here marked as TOOT, 6C, 
7C and 3C) and the high-z radio-loud samples (Hz3CR and HzLLRG) are plotted separately. 
The color code is the same as for the left panel. \label{evol}}
\end{figure*}

\subsection{Radio loud samples: no trends with redshift and luminosity}

In this section we discuss the results we obtained for the  different radio loud samples.
It is in fact particularly interesting to investigate whether the merger fraction in these object
may depend on e.g. redshift, luminosity, or on the original criteria for each sample selection.
However, while the statistical analysis clearly shows that radio loud objects are almost all associated with mergers, 
one important {\it caveat} is  that the samples are small. This is particularly true for 
the radio-loud samples (except for the low-z 3CR), but also for the high-z radio-quiet comparison sample.
Unfortunately, this prevents us from drawing statistically firm conclusions on any possible trends 
between the different samples of radio galaxies. For example, we cannot determine whether the HzLLRG's are
less likely to be associated with mergers than the Hz3C's even if the observed fraction is lower in the former 
than in the latter. Similarly, we cannot determine whether each of the samples
at $z\sim 0.5$ that compose the Willott sample behave differently, for example as a result of their different
radio power.

However, it is  interesting to consider radio-loud samples grouped by redshift bin. We can thus compare the merger fractions 
for the low-z 3CR, the whole Willott sample (TOOT+7CRS+6CE+3CRR),
and the high-z objects (3CR+HzLLRG). 
In Sect.~\ref{stats} we showed that we  find no statistical evidence that that the observed merger fractions are different for
any of those groups. Therefore, while we wish to stress that this does not imply that they are the same, we can 
conclude that the data show no evidence for a redshift evolution. By comparing the 95\% Bayesian credible intervals,
we can also conclude that the merger fractions do not differ by more than $\sim 20\%$.
Since the samples are well separated in radio luminosity, it is straightforward to perform a similar analysis for
samples grouped by radio luminosity, and conclude 
that the  merger fraction in low and high-power samples does not differ by more than 20\%.
This is a notable result, since these samples are separated by more than 4 dex in radio power, and span a wide range of redshift 
(see Fig.\ref{rlsamples}).

In Fig.\ref{evol} we show the Bayesian
95\% credible intervals for each of the groups of radio loud objects plotted against the
redshift (left panel) and radio luminosity (right panel) range spanned by  each group. 
In the left panel (merger fraction vs redshift) we also include 
the radio-quiet AGNs and the two samples of 
non-active galaxies. The radio-quiet AGNs and the non--active galaxies are not plotted in the right panel, since any trends with the radio power would be irrelevant. 
In fact, the origin of radio emission in radio-quiet AGN
is still debated, as it could be either thermal or non-thermal, and a possible contribution from starbursts could 
not be excluded at the lowest luminosities. Starbursts are instead the most likely origin for the radio emission in non-active galaxies. These
objects would lie on the bottom-left of the figure in the right panel, but any correlation with the merger fraction
would be meaningless.
As it is clear from the figure (left panel), the only group of radio loud AGNs that is still marginally compatible with the 
radio quiet samples is the Willott sample. However, this might be due to the fact that the images were taken with
WFPC2, which was significantly less sensitive than WFC3-IR. 
In Tab.~\ref{surfbrightlim} we report the limit magnitude estimated using the WFPC2 Exposure Time Calculator, and
converted to the F140W filter using {\it synphot} to allow an easier comparison with the WFC3 observations. 
The value m$_{F140w}=20.7$ mag arcsec$^{-2}$ was derived using an elliptical galaxy spectrum redshifted to z=0.5.
As a result of that, some images might not show faint 
surface brightness structures such as e.g. asymmetries or tidal tails, which might lead us to misclassify some of the objects as non-mergers.
Deeper images with WFC3 or with ACS in the I band should be taken in order to achieve a more reliable estimate of
the merger fraction in the Willott sample.

The right panel of Fig.~\ref{evol} shows the merger fraction for each sample, against radio power at 
151MHz\footnote{Measurements at radio frequencies
different form 151MHz were converted to the reference wavelength using a spectral index $\alpha = 0.8$ and 
$F_{\nu} \propto \nu^{-\alpha}$.}. Here we plot the Willott samples separately, since they belong to different
luminosity bins. The low-z 3CR span a large range in radio power. Most of these 3CR objects are confined to the lower luminosity bin, as it is 
clear from Fig.~\ref{rlsamples}. As noted above, except for the low-z 3CR for which the merger fraction is much better 
constrained, the number statistics is small and the error on the merger fraction is large. However the figure
clearly shows that all of the samples are located in the upper part up the diagram, and that no trend with 
radio power is visible.

\begin{figure}
\epsscale{1.1}
\plotone{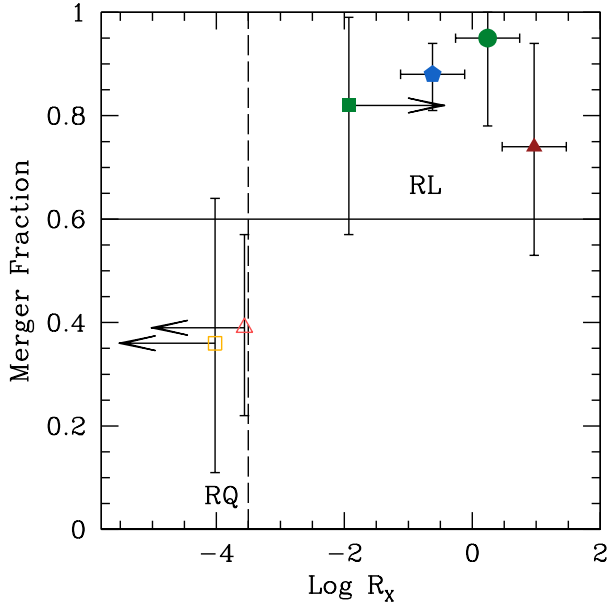}
\caption{Merger fraction vs. average radio loudness parameter R$_X$ for the different AGN samples. 
Filled symbols are the radio loud samples, empty
symbols are radio quiet. The Hz3C sample is plotted as a circle, the HzLLRG is the square (lower limit), 
Willott's 3CRR is the triangle, and the low-z 3CR is the pentagon. For the radio quiet samples (empty symbols) 
the LPTy2AGN sample is the triangle, and the HPTy2AGN is the square.
The dashed line represents the radio loudness threshold for PG QSOs \citep{terashima03}. 
The solid line marks the 60\% merger fraction 
that appears to roughly separate radio--loud and radio--quiet samples. Color code is the same as for Fig.~\ref{evol}
\label{radioloudness}}
\end{figure}

Finally, in Fig.~\ref{radioloudness} we show the merger fraction against the radio loudness parameter R$_X$. 
In order to represent the average value of R$_X$ for each sample, we
calculated average values for both the radio and the X-ray luminosities (see the caption for references).
The data are very uncertain, since 
the information is extremely sparse. This especially holds  for the X-ray data of the Hz3C and the Willott sample, while 
most of the HzLLRG only have upper limits in the X-ray band, so the radio loudness of the HzLLRGs is represented as a lower limit. 
Note that $\sim 20\%$ of the radio-quiet objects are
detected in the radio band. 
The value of R$_x$ for the radio-quiet samples is thus calculated using the average radio luminosity for the detected objects, and the
points in the figure are shown as upper limits for the radio loudness parameter. For all other samples, 
to be safe, we assume uncertainties up to $\sim \pm 0.5$dex in R$_X$. 
This is reasonable considering the uncertainties in the X-ray luminosity and the
range of radio and X-ray luminosity spanned by the objects. 
%We also tested that a linear fit cannot reproduce the data. 

This plot summarizes the main result of this work, i.e. that radio quiet sources are systematically 
associated to smaller merger fractions (and they are located in the lower-left side of the figure), while radio loud AGN are 
unambiguously associated with mergers (and they are located in the upper-right side of the figure). 
%Note that the transition between radio-quiet and
%radio loud occurs at R$_{X,t} \sim -2$ for the samples considered in this work. This value is different from the one adopted by \citet{terashima03}, 
%which found Log R$_{X,t} \sim -3.5$ for PG quasars. However, 
%here we use a slightly different definition for R$_X$, since the radio frequency is different from that
%used by \citet{terashima03}, and we use the total radio power instead of isolating the nuclear luminosity. 

\subsection{Comparison with other recent results}
Our results basically agree with the findings of \citet{ramosalmeida12} and \citet{ramosalmeida13}. 
Those studies focus on samples of relatively low-redshift
($z<0.7$) sources with deep ground based observations. 
These authors found that the large majority ($\sim 80\%$) of radio-loud objects show disturbed morphologies, and
they also reside in  dense environments. However, the parameter space covered by those papers is limited to luminous low-redshift objects, 
while our work spans a significantly larger range both in redshift and luminosity. 
The same group also studied a sample of 20 Type~2 quasars \citep[]{bessiere12} that includes at least two RLQSOs, based on their location in the 
L$_{5GHz}$ v L$_{[OIII]}$ plane \citep[]{xu99}. As expected in light of our results, both RLQSOs are associated with mergers.
The merger fraction in the full sample is as high as 75\%.   If we
only exclude those two sources that exceed their definition of radio-loudness,
the observed merger fraction among the remaining radio-quiet objects is still quite large (72\%). 
While this might appear in disagreement with our measured fraction for the RQAGNs, the uncertainty on the value found by these authors is large, 
as a result of the small number statistics (13 possible mergers out of 18 objects). 
Since the samples are small, we cannot reject the null hypothesis that a 72\% merger fraction for that sample 
is different from the fraction measured in our sample of RQAGNs.

One of the major points here is that we showed that there is no evidence for a trend with luminosity, at least for the radio loud samples. This is
apparently at odds with the results of \citet[]{treister12}, where only the most powerful samples appear to predominantly reside  in merging systems. 
However, it is interesting to briefly discuss their work in the light of our results. Two samples show a significantly greater merger 
fraction than any other samples treated in that work. Those are
the sample of dust-reddened Type~1 QSOs of \citet[]{urrutia08} and the \citet{bahcall97} sample of
Type~1 QSOs. The former is most likely biased in favor of a high merger fraction, because of the nature of those obscured quasars. The latter,
instead, is more interesting. Of the 20 QSOs observed with HST by \citet{bahcall97}, 14 are radio quiet and 6 are radio loud. All of the 6 RLQSOs
are apparently merging or show irregular features that might be explained with a merger, in agreement with our results. 
Thus if we only limit the analysis to the RQQSOs, the fraction of mergers is  $\sim 50\%$, in agreement with our findings.
Therefore, we argue that the high fraction of mergers in the full Bahcall sample (65\%) could in principle be explained, in light of our results, 
by the fact that a  significant fraction of the objects are radio loud.

\section{Implications for the origin of radio-loudness}
\label{implications}
Our results show that radio galaxies are unambiguously associated with mergers, independently of redshift
and luminosity, while radio-quiet AGN at $z>1$ are indistinguishable from normal galaxies in the same redshift range. 
This result may have profound implications for our understanding of 
the mechanisms that enable the production of powerful relativistic jets from supermassive BHs. 
In fact, such a strong connection between mergers and radio loud AGNs may be a clue for a direct link between these two phenomena. 

The central question here is whether mergers may
provide, or at least substantially contribute to  the physical conditions that ultimately enable the formation of jets in RLAGNs. 

\subsection{Not all mergers may generate a RLAGN}
First of all, we should point out that most mergers do not generate a RLAGN. This is clear from the fact that a fraction 
of non-active galaxies at $z>1$ are seen to be merging, but they show no signs of radio-loud activity. 
If there is an association between these two phenomena, it is not a univocal cause/effect relationship.
Based on our results, the same also holds for radio-quiet AGNs since a fraction of those are associated with mergers. 
Thus, we conclude that mergers are unrelated to radio-quiet AGNs or, alternatively, only a
fraction of those may be triggered by mergers. However, it is worth noting that our definition of merger includes
objects that still have to merge as well as objects for which the signs of a past merger are somehow still visible. Therefore,
the timescale during which we observe a merger is probably of the order of at least $\sim$ 1-2 Gyr \citep[e.g][]{dimatteo05}. 
The time scale for radio
loud activity is most likely significantly shorter ($\sim 10^{7} - 10^8$ yr) . Therefore, we cannot exclude that some of the 
non-active galaxies that we observe in a merger phase are turned-off radio loud AGNs, or, alternatively, 
they still have to be turned-on. 
Summarizing, we believe that not all mergers may directly generate a radio-loud AGN. Below we discuss a few possible 
conditions for a merger to trigger RLAGN activity.

\subsection{Conditions for mergers to trigger RLAGNs}
We argue that when certain  conditions are met, mergers may trigger radio loud nuclear activity. 
What we ultimately wish to know is what those conditions are.
An important piece of information here is that the association between mergers and RLAGNs is robustly established at all redshifts. 
While both recent simulations and  observations show that the galaxies merger rate increases with 
redshift \citep[e.g.][]{hopkins06,conselice03,lotz11,rodriguezgomez15}, our results
show that there is no evidence that the merger fraction for RLAGN is higher at higher $z$. 
Of course, this should be confirmed through the analysis of larger samples that may better constrain the merger fractions
and highlight any possible trends with redshift. But even if the uncertainties remain large with the present samples, 
the existence of such a tight relationship between the two phenomena means that somehow 
a RLAGN needs a merger (at all redshifts) in order to manifest itself. 
The predominance of major mergers among our radio loud samples (at least at $z>1$) that is apparent from visual inspection needs to be 
confirmed  through a more quantitative analysis. However, this may imply 
that one of the conditions that must be met in order to trigger a RLAGN is that the merger needs to be between two galaxies 
(and thus between two BHs) of similar mass. It is well known that radio galaxies are ubiquitously associated with very massive hosts 
\citep[$\gtrsim 10^{10-11}$ M$_\sun$, e.g.][]{best05} and high-mass SMBH \citep[$\gtrsim 10^8$ M$_\sun$][]{laor00,chiabmarconi, 
dunlop03,bestheckman12}. Therefore, we expect those major mergers to involve high mass objects only.

\subsection{How do mergers affect the central supermassive black holes?}
Understanding the details of this issue is central to our future work on the subject, but it is beyond the scope of
this paper. However, we can build on our results and speculate on a possible scenario.
One of the effects of mergers is to lower the specific angular momentum of the gas in the galaxy, and thus to drive the gas towards the center
\citep[e.g.][]{hernquist89,hopkinsquataert11}. While this may naturally happen in gas rich mergers, as those observed in our
high-z 3CR sample \citep[e.g.][]{barthel12,tadhunter14,podigachoski15}, the hosts of low redshift radio galaxies (in particular those of low radio powers) are often relatively gas--poor systems. 
Therefore, this effect may play a role at higher redshifts, but it is very unlikely to be the
ultimate cause of radio loud activity in general. Furthermore, tidal effects happen in all mergers, thus merger events should affect
both RQ and RL AGNs at the same level, contrary to our results.

Another effect of mergers is to alter the spin and the mass of the central black hole. That can be achieved in two ways,
i.e. either via accretion, or via BH-BH merger \citep[e.g.][]{volonteri13}. 
The former implies that a large amount of gas is driven toward the central region
of the galaxy, for a significant amount of time. If the accretion of matter is coherent, i.e. if the flow of matter 
occurs at a fixed angular momentum axis, that ultimate leads to spinning-up the black hole. 
On the other hand, if accretion results from multiple merger events that drive the matter towards the 
black hole from different directions,
the BH is spinned-down. In any case, as already pointed out above, the expected amount of gas in major mergers 
(at least at low redshifts)
is probably too small to alter the BH spin significantly in these objects.

In the  case of BH-BH merger, the two BH coalesce and the resulting object has increased mass and, in most cases, 
a lower spin per unit mass. 
But there are scenarios in which those events lead to the opposite result. For example, it has been shown that a single
major BH-BH merger, where the ratio between the masses of the two involved BHs approaches unity, may 
generate a spinning BH, even if the two merging black hole are initially not spinning 
\citep[][for a recent review]{hughesblandford,baker06,li10,giacomazzo12,schnittman13}. 
However, current simulations are unable to
reproduce BH with dimensionless spin parameter greater than $\sim 0.94$  as a result of BH-BH mergers alone \citep[]{hemberger13}.

\subsection{The role of rapidly spinning black holes}
%\citet{wilsoncolbert} \citep[and see also][]{hughesblandford} showed that 
%by assuming that major BH-BH mergers are the main cause of BH spin-up in radio-loud AGN, it is possible
%to reproduce the AGN radio luminosity functions with good accuracy. 
%However, recent work
%by \citet{volonteri13} and \citet{li12} achieve slightly different conclusions that seem to contradict the 
%{\it spin paradigm}. Based on
%the fact that since the merger rate depends on redshift, these authors point out that the fraction of radio loud objects
%should be higher at high redshifts, contrary to recent findings \citep[]{jiangfan07}.

According to the so-called {\it spin paradigm} \citep[]{wilsoncolbert,blandfordsaasfee}, 
radio loud AGNs are associated with rapidly  spinning black holes, while BHs in radio quiet AGN are expected to spin less rapidly.
Those major mergers that result in rapidly spinning black holes  may provide the 
link between our observations and the physics behind the RLAGN phenomenon as a whole.
Clearly, the objects that we are seeing in a pre-merger phase, or the very few ones that 
are associated with a minor merger, do not fit the above scheme. In those cases, we must assume that another major merger event
happened in the recent past. This does not seem unreasonable, since all of these objects lie in over-dense environments, but
it should be proven by finding the signatures of that previous merger. 

In a framework in which BHs are spun up by major BH-BH mergers, we expect a range of resulting 
spin values. Therefore, it is possible that different radio morphologies (and radio powers) are associated with different 
BH spin levels. For example, 
only the BHs that spin more rapidly might be able to produce the most relativistic jets, in a framework in which the jet
is powered by energy extracted from the rotating black hole \citep[e.g.][]{blandfordznajek, mckinney12,ggnature}. 
Such a scenario has been explored for X-ray binaries jets, so far leading to contrasting results 
\citep[e.g.][]{gardnerdone14a,narayanmcclintock12,fender10,russell13}. 
It is interesting to note that \citet{fanidakis11} were able to reproduce the RL and RQ AGN populations 
in the context of galaxy evolution. Their model is based on a scheme that includes a bimodal BH spin distribution mainly caused by
the the combination of accretion of matter onto the black hole and by the different type of mergers that galaxies of 
different stellar mass undergo.

As pointed out in the previous section, black holes could also be spun-up by merger-triggered accretion.
Such a  mechanism could in principle account for our observations of mergers at $z>1$, but only if those are gas-rich mergers. 
However, it would not completely explain the origin of radio-loud activity in low redshift, gas-poor galaxies 
(see also the discussion in the 
next Section).

\subsection{Low redshift radio loud AGNs}
A scenario in which BH-BH mergers are directly implied in triggering radio loud AGN activity seems to be supported by the
observed properties of low-redshift RLAGN  hosts.
\citet[]{sandrobarbara06,deruiter05} showed that all radio loud AGNs in their samples  are hosted
by core-galaxies, i.e. galaxies that show a flat radial brightness profile in the central regions.
While this analysis is clearly limited to very low redshifts ($z\lesssim 0.1$), where the core radius can be resolved in HST images, and to
objects that are not affected by significant amounts of dust in the central kiloparsecs, it very clearly shows that
there is a strong connection between the presence of a {\it core} profile and RL activity. 
Interestingly, one of the most likely explanations for the presence of core profiles  is related to
a  major BH-BH mergers, in which the binary BH formed during the merger ejects stars from the central regions before 
the two BHs coalesce \citep[e.g.][]{graham04,merritt06}. While the direct physical connection between such a phenomenon
and radio loud activity is still a matter of debate \citep[see e.g.][]{chiabmarconi}, it provides one more
piece of evidence that major mergers and radio loud AGNs are strictly connected.

\section{Conclusions}
\label{conclusions}
We derived the merger fraction in samples of $z>1$  Type~2 RLAGN, Type~2 RQAGN, and non- active galaxies.  
We  establish that the RLAGN are unambiguously associated with mergers (92\%$^{+8\%}_{-14\%}$), while only 38\%$^{+16}_{-15}$ 
of the RQAGN show evidence for a merger. Non-active  galaxies are statistically indistinguishable from the RQAGNs.
The comparison with lower redshift samples  shows that there is no evidence for the fraction of mergers in RLAGNs 
to be dependent on either redshift or AGN luminosity.

Mergers are directly involved in triggering radio-loud activity at all redshifts.  We speculate that the galaxy mergers
we observe at $z>1$ are responsible for spinning up central black holes possibly through mergers of high-mass BHs.

It will be extremely important to study the BH mass distribution in samples of merging galaxies of different type and to
determine whether the BH mass and the merger types are related to different type of activity. 

Not all galaxy mergers in our RL samples appear the same to visual inspection. 
It would be important to firmly assess the ratio between major and minor mergers at different redshifts, and  radio power.  
The images in this work  show that 
a larger fraction of objects are in a phase of ongoing- or post-major merger in higher redshift samples than in the 
lower redshift counterparts \citep[see also][]{floyd08}.  The different HST cameras used not only  sample different 
rest-frame wavelengths (near-IR at low-z
and optical at $z>0.4$), but they also have significantly different sensitivities. 
It is extremely hard to
detect post-merger signatures such as e.g. faint tidal tails in the intermediate redshift samples imaged with WFPC2. 
Those features may lie outside the field of view in the NICMOS images at low redshift.
The only redshift range that is covered with homogeneous
observations is between z=1 and 2.5.  
Clearly needed is a homogeneous data set observed with HST-WFC3 or ACS at all redshifts.

Detailed studies of the mergers in our radio-loud samples using integral field spectroscopy, combined with deep high 
resolution imaging in the rest frame 
IR , should be used to study the kinematics of the mergers and measure a range of parameters such as dynamical masses and angular momenta.
The imaging part of the project is feasible with WFC3, although at high redshifts the rest frame wavelengths sampled by such a camera 
are still within the optical bands. Dust obscuration may reduce the accuracy of the stellar mass measurements.

ALMA observations can trace the molecular gas involved in the mergers. It is important to study a large
sample of these sources at different redshifts, to test whether different amounts of gas are driven towards the central supermassive black holes
and whether that may affect the type of radio loud nuclear activity  A comparison with the mergers observed in radio-quiet 
AGNs and in ULIRGs will elucidate the black hole feeding mechanism in all of these sources, and relationships with  AGN activity. 

This work is limited to Type~2 AGNs because in Type~1 AGNs the
nuclear light from the central QSO hampers morphological studies of the host galaxy. With a better knowledge of the
WFC3 PSF, such a study is possible starting with the sample of 3CR QSOs at $z>1$.
Better results can be achieved using coronographic observations by \citet[]{martel03}.
While the coronographic mode is now available on HST only in STIS,
JWST with NIRISS will allow us to image  QSOs spanning a broader range of redshift.
We hope to be able to study in detail both the structure and fueling mechanisms in Type~1 objects \citep[see e.g.][]{ford14}.

In conclusion, our results clearly establish that RLAGNs are  mergers. Conversely, based on the samples studied in this paper, 
we did not find any statistical evidence that RQAGNs are related to merger events.

\section{Acknowledgments}
The authors wish to thank G. Zamorani, C. O'Dea, S. Baum and A. Capetti for insightful comments and suggestions. 
We acknowledge W.~B. Sparks and the HST-3CR collaboration for making the
images of the $z>1$ 3C radio galaxies available for this work. We are
also indebted to M. Stiavelli and A. Mortazavi for helping out with the
mergers classification. 

We are also greatful to the anonymous referee for his/her helpful comments that greatly improved the paper.

This research has made use  of the NASA/IPAC Infrared Science Archive,
which  is  operated  by  the  Jet  Propulsion  Laboratory,  California
Institute of Technology, under  contract with the National Aeronautics
and Space Administration.  This work is based on observations taken by
the CANDELS Multi-Cycle Treasury  Program with the NASA/ESA HST, which
is  operated  by  the  Association  of Universities  for  Research  in
Astronomy, Inc., under NASA contract NAS5-26555. This work is based on 
observations taken by the 3D-HST Treasury Program (GO 12177 and 12328) 
with the NASA/ESA HST.

\facility{Based on observations made with the NASA/ESA Hubble Space Telescope, 
obtained from the Data Archive at the Space Telescope Science Institute, which is operated by the Association 
of Universities for Research in Astronomy, Inc., under NASA contract NAS 5-26555.}

%\bibliography{mybiblio}

\end{document}